\documentclass[fleqn,usenatbib]{mnras}

\usepackage{newtxtext,newtxmath}

\usepackage[T1]{fontenc}

\DeclareRobustCommand{\VAN}[3]{#2}
\let\VANthebibliography\thebibliography
\def\thebibliography{\DeclareRobustCommand{\VAN}[3]{##3}\VANthebibliography}

\usepackage{graphicx}	
\usepackage{amsmath}	

\usepackage[utf8]{inputenc}
\usepackage[T1]{fontenc}

\title[Stellar populations of very metal-poor extragalactic globular clusters]{Horizontal branch structure, age, and chemical composition for very metal-poor extragalactic globular clusters}

\author[M. E. Sharina et al.]{
M. E. Sharina,$^{1}$\thanks{E-mail: sme@sao.ru}
M. I. Maricheva,$^{1}$
A. Y. Kniazev,$^{2,3,4,1}$
V. V. Shimansky,$^{1}$
I. A. Acharova$^{5}$
\\
$^{1}$Special Astrophysical Observatory, Russian Academy of Sciences, Nizhnii Arkhyz, 369167 Russia\\
$^{2}$South African Astronomical Observatory, PO Box 9, 7935 Observatory, Cape Town, South Africa \\
$^{3}$Southern African Large Telescope Foundation, PO Box 9, 7935 Observatory, Cape Town, South Africa \\
$^{4}$Sternberg Astronomical Institute, Lomonosov Moscow State University, Moscow, Russia\\
$^{5}$Department of Physics, Southern Federal University, 5 Zorge, Rostov-on-Don 344090 Russia\\
}

\date{Accepted XXX. Received YYY; in original form ZZZ}

\pubyear{2023}

\begin{document}
\label{firstpage}
\pagerange{\pageref{firstpage}--\pageref{lastpage}}
\maketitle

\begin{abstract}
This paper presents the results of analysing the integrated light (IL) low-resolution spectra of globular clusters (GCs) in the M31 and Centaurus~A groups of galaxies. The sample consists of eight very metal-poor GCs ($\rm [Fe/H]\le -2$~dex) with high signal-to-noise ratio spectra acquired with the telescopes: the 6-m SAO RAS (BTA), the Southern African Large (SALT) and the 6.5-m Magellan (MMT). We study the influence of contribution of the horizontal branch stars on the hydrogen Balmer line profiles in the IL spectra. By modelling the Balmer lines, as well as the metal lines in the observed spectra, we determine the optimum parameters of stellar evolution isochrones and, consequently, the parameters of the atmospheres of the cluster stars. 
For all the studied GCs, the parameters of horizontal branch stars set by the selected isochrones, the corresponding ages, and carbon abundances are presented for the first time. 
The abundances of several other elements (Mg, Ca, Ti, Cr, and Mn) were determined for five GCs for the first time. 
All the studied GCs have blue horizontal branches and are older than 10~Gyr. Their chemical abundances, with the exception of Mg and Mn, are in good agreement with the abundances of stars in the Galactic field. The reasons of low [Mg/Fe] and of high [Mn/Fe] are discussed. Study of the fundamental properties of stellar populations in old globular clusters facilitates a better understanding of the formation processes of their parent galaxies and nucleosynthesis in the early Universe.

\end{abstract}

\begin{keywords}
star clusters: general – star clusters: individual: [VFH2013]~PA-N147-1, [H76b]~3, [CS82]~C39, [H32]~VIII, Bol~2, [H32]~74, EXT~8, [GPH2009]~KK~197-2
 -- galaxies: abundances -- galaxies: dwarf -- galaxies: individual: M31, M33, NGC205, NGC147, [KK98a] 197.
\end{keywords}


\section{Introduction}
\label{Intro}
Analysis of the integrated light (IL) spectra of star clusters is one of the effective tools for determining their age and metallicity and studying the formation and evolution of their host galaxies \citep[see, for example,][and references therein]{Lee23, CabreraZiri22, L22, Leath22, Fan20}. It has been shown in a number of studies that  determination of the age of old globular clusters (GCs) from the IL spectra is associated with the problem of taking into account the radiation of the horizontal branch (HB) stars corresponding to the core helium burning phase of stellar evolution. If one does not include the IL spectra of HB stars in the analysis, then the uncertainty in estimating the age of old GCs greatly increases, since the contribution to the spectra of stars with higher helium mass fraction (Y) populating the bluer parts of HBs imitates younger age \citep{Ko08, CabreraZiri22, Leath22, Da05}. To take into account the radiation of HB stars, for example, the Balmer absorption line indices were studied \citep{Sch04, Leath22, Percival11}, the flux contribution of hot HB stars was added to standard isochrones \citep{CabreraZiri22, Ko08}, or isochrones including HBs were used \citet{Sh20} (hereafter,\defcitealias{Sh20}{S20}\citetalias{Sh20} and references therein).
\begin{table*}
	\centering
	\caption{Literature and our (marked with superscript 0) characteristics of the eight GCs in M31: (1, 2, 3) the name of the GC, the designation of it used in this article and the name of the host galaxy; (4) the right ascension and declination of the object for the epoch J2000.0; (5, 6) the absolute magnitude and radial velocity; (7, 8, 9) the age in Gyr, Z, and Y of the isochrone that we used in modelling the cluster spectrum.  
For all the GCs, except for EXT8, the isochrones by \citet{B08} were used. For EXT8, the best solution was provided by the canonical scaled-solar isochrone by \citet{P04}. In the second lines of columns (5, 6, and 7), the literature data are given for each object from the papers: \citet{L22} (L22), \citet{L21} (L21), \citet{L20} (L20), \citet{Sh06} (S06), \citet{C11} (C11), \citet{G09} (G09), \citet{F20} (F20), \citet{H14} (H14), \citet{V13} (V13), \citet{Ma13} (Ma13) .}
\begin{tabular}{lcccccr@{$\,\,\,$}rr}     
\hline \hline                        
Name       & Symbol for& Host   & $\alpha$(J2000) & M$_V$          &  $\rm cz$         &     \multicolumn{3}{c}{Isochrone}             \\   \cline{7-9}                      
           & object    & galaxy & $\delta$(J2000) & (mag)          &($\rm km \cdot s^{-1}$)&  $\rm Age (Gyr)$ & $\rm Z^{0}$ &  $\rm Y^{0}$        \\ \hline
   1       &   2    &   3    &  4              &   5               &  6                &   7              &  8          &  9               \\                   
 \hline                                                                                                                                                           
[VFH2013]  &  PA    & NGC147 & 00:32:35.3      &                   & -217$\rm \pm8^{0}$    & 10.0$\rm \pm0.28^{0}$& 0.0001$\rm \pm0.00001$& 0.26$\rm \pm0.005$ \\  
PA-N147-1  &        &        & +48:19:48.0     & -7.8$^{\rm V13}$  & -215$\rm \pm10^{V13}$ &  13$\rm^{L22}$      &                         &    \\   \noalign{\smallskip}
[H76b]~3   &  HIII  & NGC147 & 00:33:15.2      &                   & -219$\rm \pm12^{0}$   & 12.6$\rm \pm0.21^{0}$& 0.0001$\rm \pm0.00002$& 0.30$\rm \pm0.005$ \\       
= Hodge~III&        &        &  +48:27:23      & -8.2$^{\rm V13}$  & -197$\rm^{L22}$      &  13$\rm^{L22}$   &                             &  \\   \noalign{\smallskip}      
$\rm[CS82]$~C39&  C39& M33   &  01:34:49.6     &                  & -221$\rm \pm10^{0}$   & 12.6$\rm \pm0.07^{0}$& 0.0001$\rm \pm0.00001$& 0.30$\rm \pm0.01$   \\                            
           &        &        &  +30:21:56      & -8.6$^{\rm Ma13}$ &                   &                  &                              & \\   \noalign{\smallskip}      
[H32]~VIII &  B317  & NGC205 & 00:39:55.3      &                   & -176$\rm \pm8^{0}$    & 12.6$\rm \pm0.15^{0}$& 0.0004$\rm \pm0.00004$& 0.26$\rm \pm0.0024$ \\     
=Bol~317   &        &        & +41:47:46.0     & -8.0$^{\rm G04}$  & -178$\rm \pm28^{C11}$ & 10.0$\rm^{S06}_{CMD}$&                       &  \\   \noalign{\smallskip}       
Bol~2      &  B2    & M31    & 00:40:02.6      &                   &  -332$\rm \pm11^{0}$  & 10.0$\rm \pm0.23^{0}$& 0.0001$\rm \pm0.00001$ & 0.26$\rm \pm0.003$ \\
           &        &        & +41:11:53.5     & -7.4$^{\rm G04}$  & -338$\rm \pm14^{C11}$ &  (14)$\rm^{C11}$    &                            & \\   \noalign{\smallskip} 
[H32]~74   &  B165  &  M31   & 00:43:18.2      &                   & -67$\rm \pm7^{0}$     & 12.6$\rm \pm0.2^{0}$ & 0.0004$\rm \pm0.00004$ &  0.26$\rm \pm0.005$ \\ 
=Bol~165   &        &        & +41:10:54.7     & -8.1$^{\rm G04}$  & -68$\rm \pm14^{C11}$  &  (14)$\rm^{C11}$ &                            &  \\   \noalign{\smallskip}
EXT8       &  EXT8  & M31    & 00:53:14.5      &                   & -201$\rm \pm13^{0}$   &  11.0$\rm \pm0.3^{0}$& 0.00001$\rm \pm0.000004$& 0.245$\rm \pm0.005$ \\                   
           &        &        & +41:33:25.0     & -9.3$^{\rm H14}$  & -204$^{L20}$          &  13$\rm^{L21}$      &             &              \\   \noalign{\smallskip}
$\rm[GPH2009]$& KK  & NGC5128& 13:22:02.0      &                   & 636$\rm \pm4^{0}$     & 13.6$\rm \pm0.47^{0}$& 0.00013$\rm \pm0.00003$ & 0.26$^{+0.04}_{-0.01}$ \\ 
KK~197-2     &      &        & -42:32:08.1     & -9.8$^{\rm G09}$  & 635$\rm \pm2^{F20}$   &  6.5$\rm^{F20}$     &                        & \\   \noalign{\smallskip}
\hline \hline
 \end{tabular}
 \label{tab:2}
\end{table*}
The results of building the model IL spectra of GCs using the synthetic or observed spectra of stars and the results of analysing the distribution of stars in the colour-magnitude diagram (CMD) depend on the stellar evolution models used. However, it is not yet possible to take into account variations of some parameters during stellar evolution, for example, the variety of the convective overshoot conditions in stars depending on their chemical composition, mass, and evolutionary stage \citep[and references therein]{Viani18, Da18}. Star clusters of the same age and metallicity can have different HBs with a predominance of blue or red stars, which may occur due to differences in Y and in the mass-loss efficiency of red giants \citep[see, e.g.,][]{Tailo20, Da05, Percival11, Lee00}. The result of studying the IL spectra of GCs also depends on the quality of the observed data. Often it is difficult to take into consideration the level of the foreground and background contamination and stochastic variations in the number of stars at different evolutionary stages within the studied aperture. 

In this paper, we study the IL spectra of the representatives of very metal-poor extragalactic GCs ($\rm [Fe/H]\le -2$~dex). GCs with such metallicity are few in our and other galaxies and predominantly have blue HBs  \citep{Harris10, Beasley19}. Our analysis of low-resolution spectra ($\lambda/\Delta\lambda \sim 1000$) of GCs in nearby galactic groups is based on the method and results by \citetalias{Sh20} for the Galactic GCs in a wide metallicity range. The present study further improves this method. 
We study the influence of the properties of HB stars (T$_\text{eff}$ and luminosity) on Balmer line profiles in the IL spectra of GCs and introduce an automatic procedure of selecting the isochrones of stellar evolution for IL spectra modelling. In order to study the HB morphology, we separately consider two parameters: depth and width at half intensity (FWHM) of the Balmer lines in the IL spectra of GCs. Comparing the results of the analysis of the IL spectra of the Galactic GCs by \citetalias{Sh20} with the properties of stars on the CMD of these objects, we determine the accuracy of estimating the effective temperature T$_\text{eff}$ and luminosity of HB stars from the IL spectra. Our study will answer the question whether all the sample very metal-poor extragalactic GCs have blue HBs as their Galactic counterparts.

\section{Sample of globular clusters}
\label{sec:sample}

The sample of objects is built based on the results of our observations at the 6-m telescope of the Russian Academy of Sciences (hereafter: BTA) in 2020 - 2021 and at the Southern African Large Telescope (SALT) in 2019 - 2020. We also use the spectra acquired by \citet{C11} and references therein) during their observations with the Hectospec spectrograph of the 6.5-m MMT telescope \citep{F05}.
\citet{C09, C11} determined the radial velocities, ages, and metallicities for these and many other clusters in M31 from Hectospec observations by measuring the Lick absorption indices and comparing them with the modelled values. We selected only old GCs with $\rm [Fe/H]\le -2$~dex from the sample of \citet{C11} and analysed the spectra with high signal-to-noise ratio (SNR). According to our experience, the spectra with $\rm SNR\ge100$ per pixel in the resulting one-dimensional spectrum at a wavelength of 4500~\AA\ provide accurate results. The analysis of spectra of higher metallicity GCs will be presented elsewhere.
In the columns 4 - 6 of Table~\ref{tab:2} main characteristics from the literature for the studied objects are presented: equatorial coordinates, absolute magnitudes in the $\rm V$ band of the Johnson-Cousins photometric system and radial velocities. In the second column of Table~\ref{tab:2}, symbolic designations for the objects' names used throughout this paper are given.
The columns 7 - 9 contain parameters of evolutionary isochrones selected by the analysis of the IL spectra.
 The  IL spectra and their reduction will be characterised in Sec.~\ref{sec:2}. The methods of the spectra analysis will be described in Sec.~\ref{sec:3}. 

 All the sample GCs inhabit halos of the central massive group galaxies: M31 and NGC5128 \citep[][and references therein]{V13, Ma13, G04, L20, G09}. KK~197-2 is a nuclear GC \citep{G09} in the dwarf spheroidal galaxy KK~197 \citep{KK98}. This small galaxy contains only old stellar populations \citep{F20} and resides at $\sim$60~kpc from the centre of NGC5128 \citep{ikar07}. This is the distance, where most distant Galactic GCs reside. 

\begin{table}
\caption {Journal of spectroscopic observations at SALT (GC KK197-2) and BTA.} 
\label{jornal}
\medskip
\begin{tabular}{lcccc}
\hline \hline
 Object & Date       & T$_{exp}$        & SNR       & Seeing \\
        &            & (sec.)           & (4500\AA) & \arcsec       \\
 \hline
SALT:   &            &                 &              &         \\
KK       & 03.07.2019    & 3$\times$970  &  56       & 1.7         \\
          & 04.08.2019    & 3$\times$970  &  54       & 1.7         \\
          & 19.05.2020    & 2$\times$1000 &  43       & 1.9         \\
          & 14.06.2020    & 2$\times$1000 &  45       & 1.9         \\ \hline
BTA:      &            &                 &              &         \\
PA & 19.09.2020  & 5$\times$900    &   58      &      1.5    \\   
          & 12.11.2020  & 3$\times$900    &   51      &      2.0    \\
          & 01.10.2022  & 1$\times$900    &   31      &      1.7    \\
          & 23.01.2023  & 2$\times$900    &   39      &      1.8    \\
HIII & 19.09.2020  & 4$\times$900    &  58       &  1.8        \\
          & 12.11.2020   & 3$\times$600   &  47       &  2.0        \\
          & 04.09.2021   & 4$\times$900+  &  53       &  5.0        \\
          &            & +2$\times$600  &           &             \\
          & 06.09.2021   & 2$\times$1200+ &  48       &  3.7        \\
          &            & +600           &           &             \\
C39     & 19.09.2020 & 3$\times$600  &  64       &  1.8        \\
          & 06.09.2021    & 4$\times$1200 &  100      &  3.0        \\
 B317    & 13.11.2020    & 4$\times$600  &  79       & 1.0         \\
 EXT8     & 12.11.2020    & 3$\times$600  &  110      & 2.0         \\ 
\hline \hline
\end{tabular}
\label{tab:1}
\end{table}
\begin{table}
	\centering
	\caption{Archive data from the Hectospec spectrograph of the MMT telescope (\citet{F05}).}
 \begin{tabular}{lcccc}
\hline
 Object & Data         & T$_{exp}$ & SNR        &  ID  \\
        &              & (s)       & (4500\AA)  &  proposal   \\  \hline                                                  
B317 & 21.10.2007   & 4800      &  46        & 2007c-SAO-3  \\
        & 06.07.2005   & 3600      &  60        & 2005b-SAO-1  \\ 
B2    & 23.10.2006   & 4800      & 48         & 2006c-SAO-11 \\
        & 17.10.2006   & 2400      & 15         & 2006c-SAO-11 \\
        & 20.10.2006   & 2400      & 36         & 2006c-SAO-11 \\
        & 27.10.2005   & 3600      & 23         & 2005c-SAO-4  \\
        & 21.10.2007   & 4800      & 36         & 2007c-SAO-3  \\
B165  & 11.11.2004   &  4800     &  80        & 2004c-SAO-9   \\
        & 19.11.2006   &   4800    &  85        &  2006c-SAO-11 \\
        & 20.11.2007   &   4800    &  65        &  2007c-SAO-3  \\
\hline
\end{tabular}
	\label{tab:5}
\end{table}

\section{Observations and Data Reduction}
\label{sec:2}

\subsection{BTA and SALT data}
Observations of the sample GCs were carried out with  BTA and with SALT \citep{Buckley06, ODonoghue06}. 

The SCORPIO-I focal reducer \citep{A05} mounted at BTA was used in the long-slit spectroscopy mode. Equipped with the VPHG1200B grism, the reducer provides the spectral range of $3600-5400$~\AA\ and the spectral resolution FWHM $\sim$ 5.5~\AA\ (full width at half-maximum) with a slit width of $1\arcsec$.

We used the Robert Stobie Spectrograph (RSS) \citep{Burgh03, Kobulnicky03} at SALT in the long-slit mode with a slit width 1.25\arcsec, the grating pg0900, blocking filter pc03400 and the Camera Station parameter 26.5. The spectra cover a spectral range of 3700--6700~\AA\ with a reciprocal dispersion of 0.97 \AA\ pixel$^{-1}$
and spectral resolution of $\rm FWHM=5$\AA.  
The log of observations is presented in Table~\ref{tab:1}. The table columns for each cluster observed on a particular date show: (2) the date of observation, (3) the exposure time, (4) the SNR per pixel at a wavelength of 4500~\AA\ obtained after reduction of the one-dimensional spectra, and (5) the FWHM of stellar images. 

\begin{figure}
\centering
         \includegraphics[width=0.3\textwidth,angle =270]{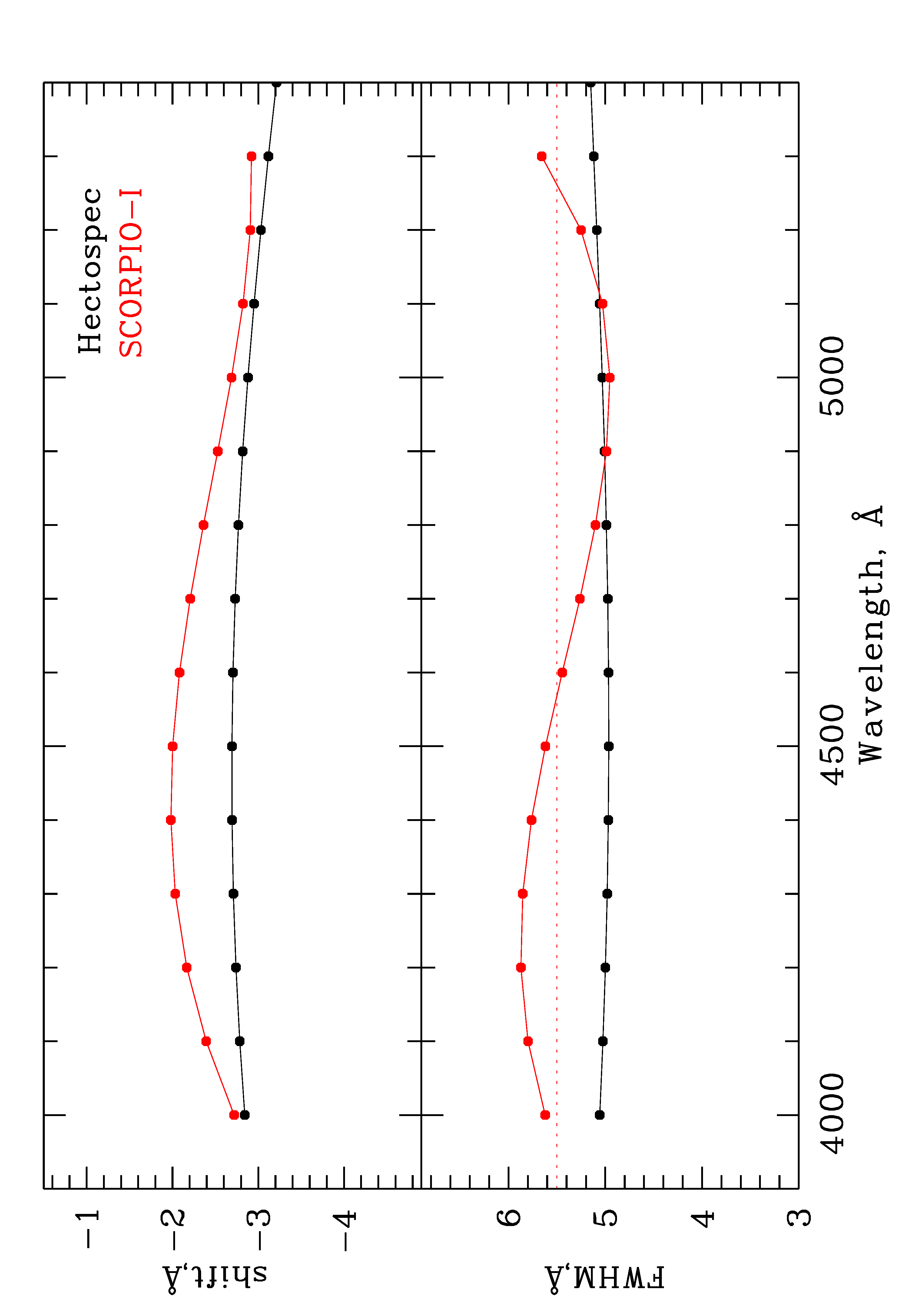}    
\caption{Variation of the measured shifts and broadening of the spectral lines in the twilight spectrum as a function of wavelength in observations with the SCORPIO-I and Hectospec spectrographs.}
    \label{fig:11}
\end{figure}
 \begin{figure*}
      \centering
        \includegraphics[width=0.49\textwidth]{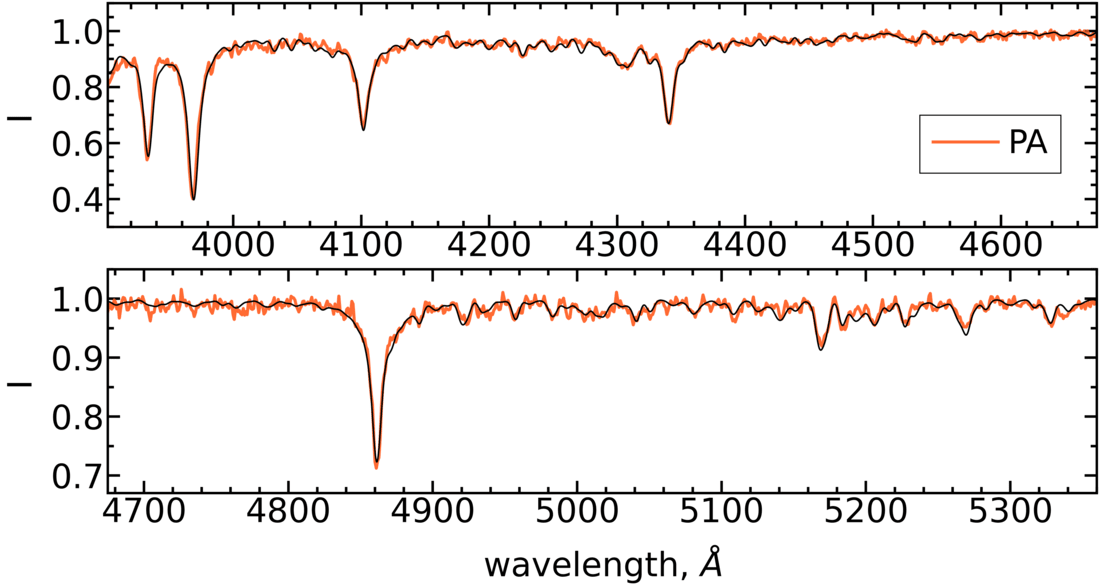}
        \includegraphics[width=0.49\textwidth]{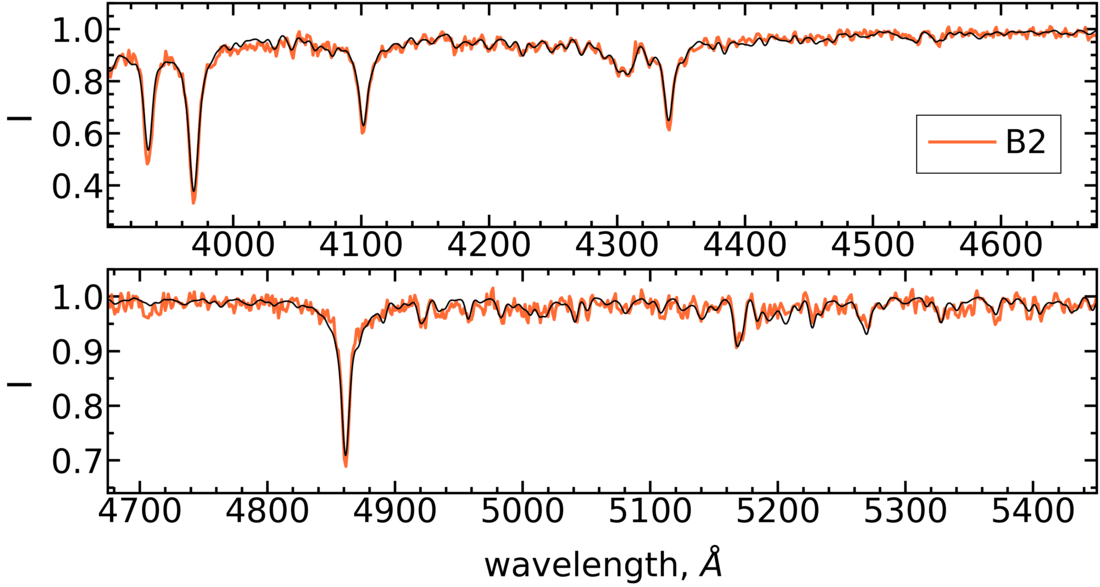}
        \includegraphics[width=0.49\textwidth]{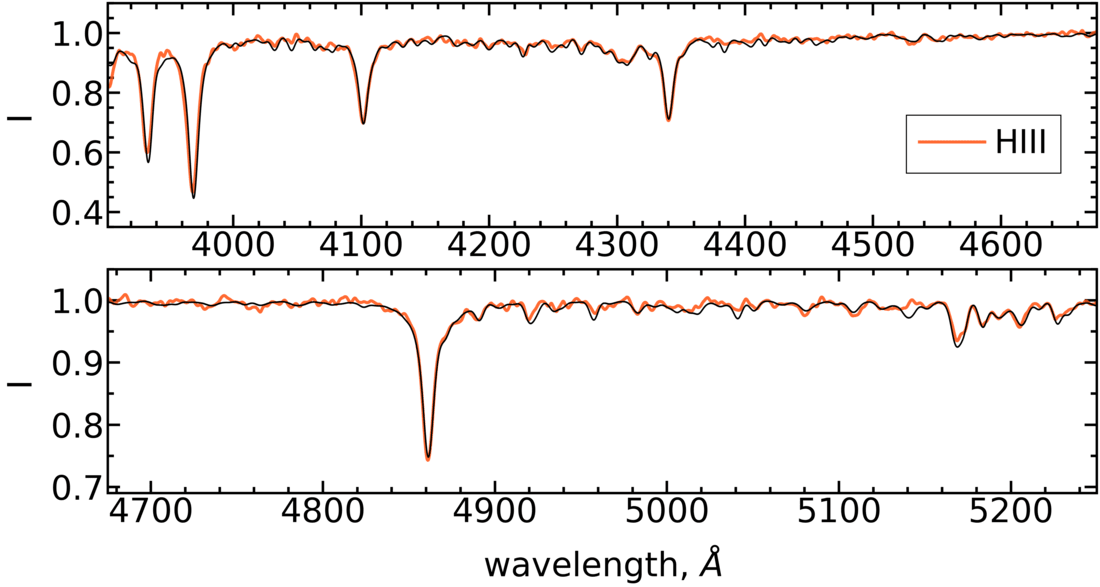}
        \includegraphics[width=0.49\textwidth]{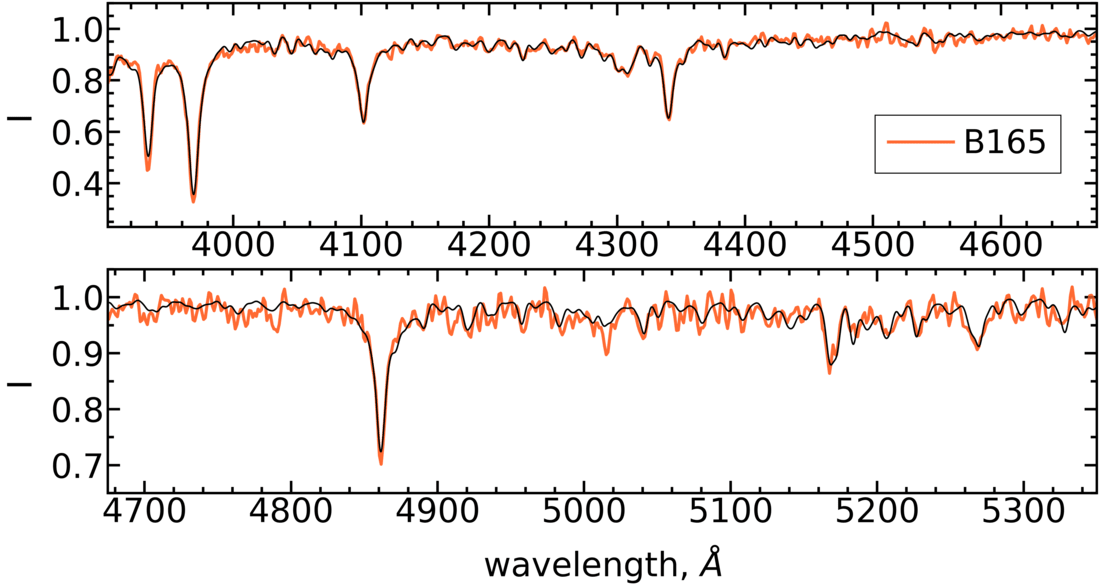}
        \includegraphics[width=0.49\textwidth]{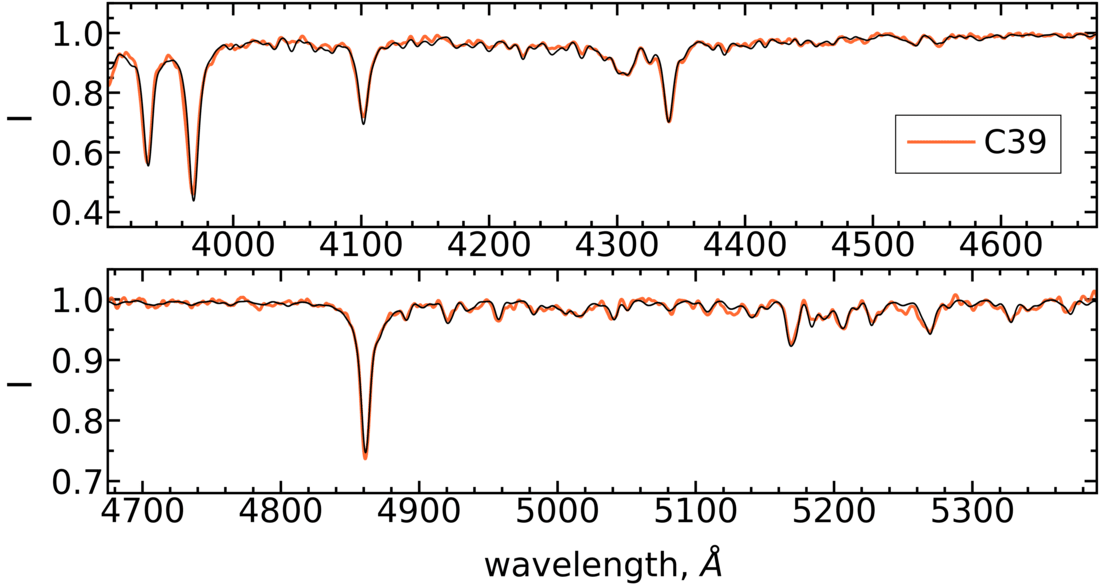}    
        \includegraphics[width=0.49\textwidth]{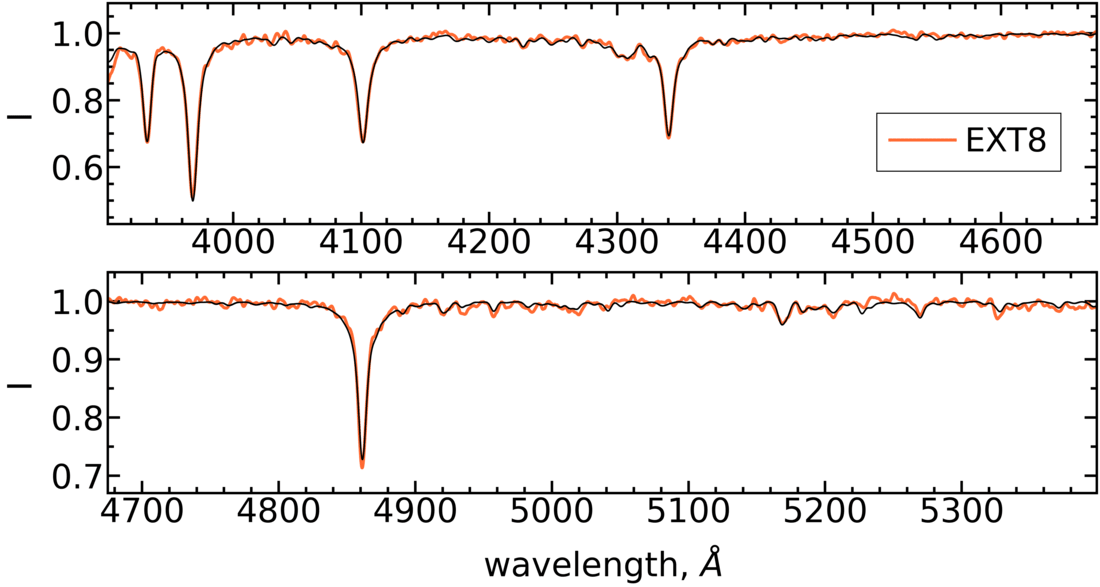}
        \includegraphics[width=0.49\textwidth]{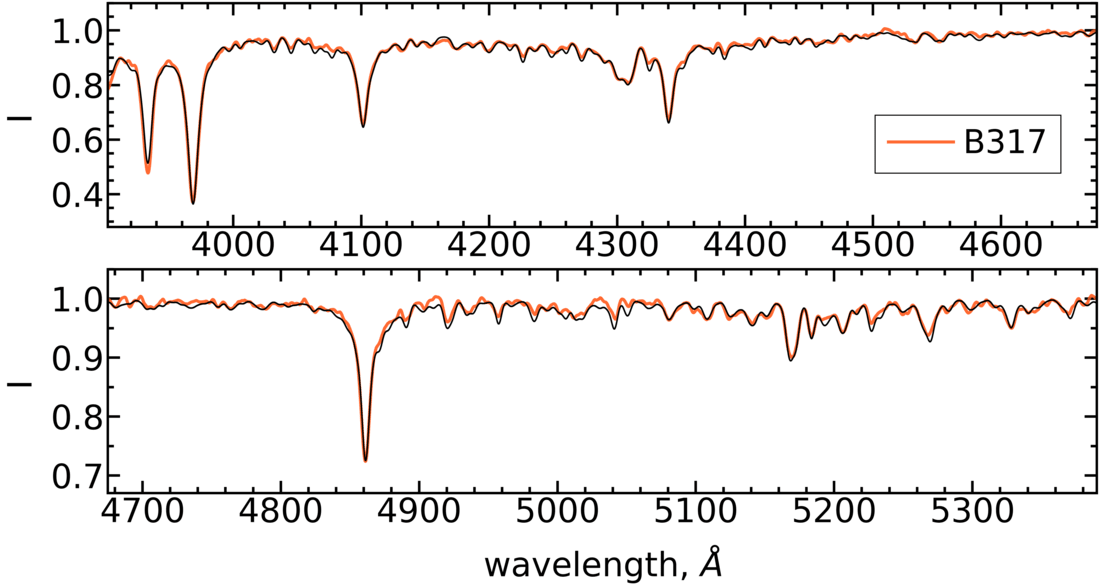}  
        \includegraphics[width=0.49\textwidth]{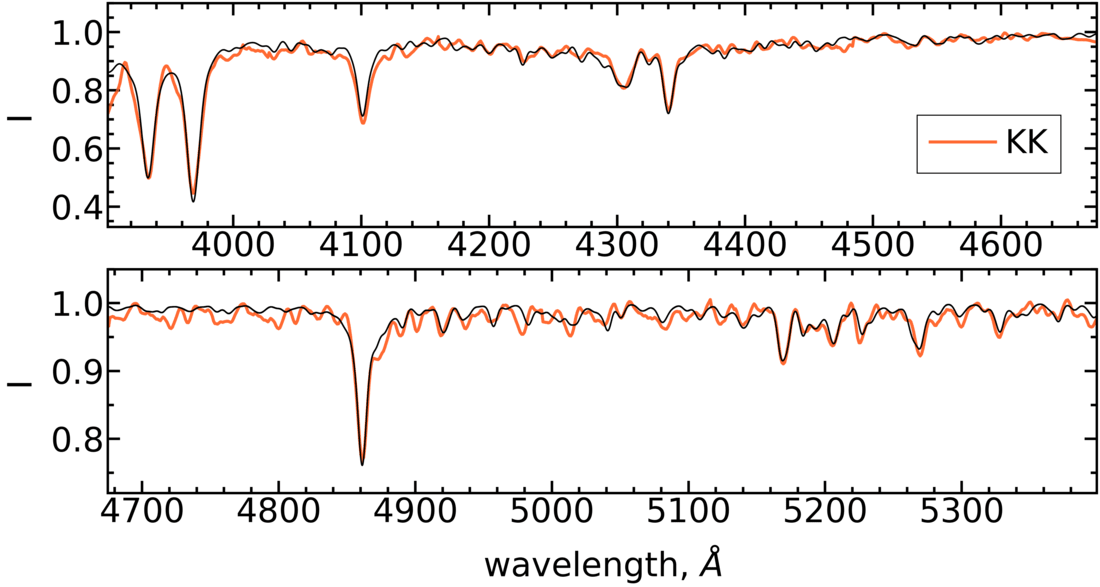}
    \caption{Comparison of the spectra of the studied clusters with the model ones calculated with the methods described in Section~\ref{sec:3}, with the isochrone parameters given in Table~\ref{tab:2}, and the element abundances presented in Table~\ref{tab:3}.}
    \label{fig:1}
\end{figure*}
The long-slit spectra were reduced using the MIDAS \citep{midas} and IRAF \citep{iraf} software packages. The primary reduction of CCD images, the cosmic ray removal, and the linearisation of the two-dimensional spectra using the successively obtained spectra of the He-Ne-Ar lamp were performed in MIDAS. The dispersion relations constructed for each object provided a wavelength calibration accuracy of about 0.16~\AA. The sky background subtraction was performed in IRAF using the \textit{background} procedure. The one-dimensional spectra were extracted in IRAF using the \textit{apsum} procedure with correction of the spectrum curvature along the dispersion.

\subsection{The 6.5-m MMT telescope data}
 The used MMT spectra of GCs B317, B2, and B165 from the observation archive of the Hectospec spectrograph of the 6.5-m MMT telescope \citep{F05}. The spectra of the clusters are given in the Hectospec spectrograph archive after reducing. They were obtained with a grid of 270~lines/mm with a dispersion of 1.21~\AA/pixel in the spectral range of $3650-9200$~\AA.  In the columns of Table~\ref{tab:5}, for each cluster observed on a specific date, the following archive data from the Hectospec spectrograph of the 6.5-m MMT telescope \citep{F05} are given: (2) the date of observations, (3) the total exposure time, (4) the SNR per pixel in the reduced one-dimensional spectrum, and (5) the program identifier. We co-added the reduced BTA one-dimensional spectrum of B317 in the spectral range of $3900-5400$~\AA\ to the corresponding MMT spectra Tab.\ref{tab:5}. 
The spectral resolution is FWHM $\sim$ 5~\AA. Before co-adding, the Hectospec spectra were normalised to a wavelength resolution similar to that of SCORPIO-I. The dependence of the spectral resolution on the wavelength and the subsequent smoothing of the spectra to the required resolution were determined based on the study of the line spread function (LSF) of the spectrograph using the \textit{UlySS} software package of the University of Lyon \citep{Ko08, Ko09}. The \textit{UlySS} web page gives examples of building the LSF: \url{http://ulyss.univ-lyon1.fr/tuto_base.html}. 
  The spectrum of any object can be represented by a convolution of the spectrum belonging to the object itself and the LSF describing the properties of the instrument. The LSF can be determined using the spectra of: 1) the twilight sky, 2) the calibration lamp in a spectrograph, or 3) the compact object (a standard star, a globular cluster). When observing the extended objects (for example, galaxies) occupying most of the CCD frame, the first or second way of building the LSF is used. Since the globular clusters under study at a distance of M31 can be seen as compact objects, it is possible to use their spectra to build the LSF. This LSF will describe the instrument function for that particular part of the CCD frame. Examples of LSFs built for the spectra obtained with the SCORPIO-I (BTA) and Hectospec (MMT) spectrographs are shown in Fig.~\ref{fig:11}. 

To measure the radial velocities of GCs (Table~\ref{tab:2}) from their IL spectra, we used the \textit{UlySS} package with the PEGASE-HR model grids \citep{LeBorgne04}, the Salpeter IMF \citep{Salpeter55}, and the ELODIE \citep{Prugniel01} stellar library. We used the instrumental LSFs built with \textit{UlySS} to determine the radial velocities.

The resulting one-dimensional spectra of the clusters are shown in Fig.~\ref{fig:1} compared with the synthetic ones calculated by the methods from the present paper and described in Section~\ref{sec:3} with selected isochrone parameters and element abundances given in Table~\ref{tab:2} and Table~\ref{tab:3}, respectively. The normalisation of the observed spectra to the synthetic ones was performed with the methods given in Appendix~\ref{app:2} that describes our procedure for the automatic selection of an isochrone for the optimum description of the observed IL spectrum. 

\section{Summarising the method of IL spectra analysis}
\label{sec:3}
 The procedure of our analysis of the IL spectra of GCs is as follows (see also \citetalias{Sh20}). Age, Y, and the approximate value of [Fe/H], determined from the isochrone parameter Z, are selected, first, using the isochrone fitting procedure described below in Sec.~\ref{sec:IsoSel} and Appendix~\ref{app:2}. Then the chemical composition including the iron abundance is specified by fitting the intensities of individual absorption features in the IL spectrum using the calculation of the synthetic spectra of GCs, that is, in fact, the chemical composition of GCs is compared with the solar one. The abundances of chemical elements are determined by varying them until the best agreement is achieved between the observed and model profiles of the spectral feature associated with a particular element. This scheme may be far from ideal but it works for the purposes of our study, as have been shown by the tests using the spectra of the Galactic GCs described in this and previous papers \citep{Sh14, Sh17, Sh20}.  Note that in the analysis of IL spectra of extragalactic GCs, we use only the comparison of the observed and synthetic IL spectra. However, in order to evaluate the results of our analysis and continue to improve our method, it is also important for us to compare the isochrones selected with our method for the specific GCs and the observed CMDs, if any. In the following, we will depict the details of the method more specifically.

\begin{table*}
	\centering
	\caption{The first line for each object shows the LTE abundances of chemical elements obtained by us for seven GCs in M31 from the modelling of their IL spectra (marked with superscript $^{0}$). The second line shows the literature LTE abundances for each object, if any.  Superscripts L22, C11, and F20 mean references to the papers by \citet{L22}, \citet{Col11} and \citet{F20}, accordingly. }
\begin{tabular}{lllllllc}
\hline 
Parameter/    &  $\rm [Fe/H]$        &  $\rm [C/Fe]$     &  $\rm [Mg/Fe]$      & $\rm [Ca/Fe]$       & $\rm [Ti/Fe]$       & $\rm [Cr/Fe]$       &  $\rm [Mn/Fe]$           \\
 Object       &    (dex)             &  (dex)            &  (dex)              & (dex)               & (dex)               & (dex)               &  (dex)               \\ 
\hline                                                        
PA            & -2.1$\rm \pm0.16^{0}$    & -0.40$\rm \pm0.12^{0}$& -0.10$\rm \pm0.13^{0}$  & 0.20$\rm \pm0.08^{0}$   &-0.10$\rm \pm0.20^{0} $  & -0.30$\rm \pm0.2^{0}  $&  0.00$\rm \pm0.2^{0}   $ \\ 
              & -2.3$\rm \pm0.02^{L22}$  &                   &  0.23$\rm \pm0.1^{L22}$ & 0.22$\rm \pm0.05^{L22}$ & 0.32$\rm \pm0.09^{L22}$ & 0.02$\rm \pm0.88^{L22}$ & -0.05$\rm \pm0.13^{L22}$   \\ \noalign{\smallskip}                   
HIII          & -2.25$\rm \pm0.15^{0}$   & -0.30$\rm \pm0.15^{0}$& -0.10$\rm \pm0.15^{0}$  & 0.20$\rm \pm0.11^{0}$   &-0.10$\rm \pm0.19^{0}$   & -0.1$\rm \pm0.2^{0}$   & -0.1$\rm \pm0.21^{0}$     \\   
              & -2.36$\rm \pm0.04^{L22}$ &                   &  0.13$\rm \pm0.1^{L22}$ & 0.35$\rm \pm0.05^{L22}$ & 0.29$\rm \pm0.08^{L22}$ & 0.18$\rm \pm0.1^{L22}$  &  0.17$\rm \pm0.13^{L22}$ \\ \noalign{\smallskip} 
C39           & -2.2$\rm \pm0.15^{0}$    & -0.07$\rm \pm0.13^{0}$& -0.15$\rm \pm0.13^{0}$  & 0.20$\rm \pm0.09^{0}$   & 0.30$\rm \pm0.18^{0}$   & 0.1$\rm \pm0.17^{0}$    &  -0.4$\rm \pm0.19^{0}$    \\  
              &                      &                   &                     &                     &                     &                     &                      \\  \noalign{\smallskip}
B317          & -2.02$\rm \pm0.16^{0}$   &  0.02$\rm \pm0.17^{0}$&  0.27$\rm \pm0.17^{0}$  & 0.30$\rm \pm0.13^{0}$   & -0.10$\rm \pm0.21^{0}$  & -0.05$\rm \pm0.22^{0}$  & -0.01$\rm \pm0.22^{0}$    \\  
              & -2.1$\rm \pm0.2^{C11}$   &                   &                     &                     &                     &                     &                      \\  \noalign{\smallskip} 
B2            & -2.0$\rm \pm0.14^{0}$    & -0.15$\rm \pm0.1^{0}$ & -0.4$\rm \pm0.17^{0}$   & 0.10$\rm \pm0.15^{0}$   & 0.10$\rm \pm0.2^{0}$    & -0.30$\rm \pm0.4^{0}$   & -0.20$\rm \pm0.3^{0}$     \\
              & -2.2$\rm \pm0.2^{C11}$   &                   &                     &                     &                     &                     &                      \\  \noalign{\smallskip}
B165          & -1.9$\rm \pm0.12^{0}$    & -0.40$\rm \pm0.1^{0}$ & 0.20$\rm \pm0.12^{0}$   & 0.10$\rm \pm0.09^{0}$   & 0.10$\rm \pm 0.18^{0}$  & 0.00$\rm \pm 0.2^{0}$   &  0.00$\rm \pm0.2^{0}$   \\ 
              & -2.0$\rm \pm0.20^{C11}$  &                   &                     &                     &                     &                     &                      \\ \noalign{\smallskip} 
EXT8          & -2.8$\rm \pm0.15^{0}$    & 0.00$\rm \pm0.14^{0}$ & -0.40$\rm \pm0.17^{0}$  & 0.35$\rm \pm0.09^{0}$   & 0.30$\rm \pm0.18^{0}$   & -0.2$\rm \pm0.19^{0}$   &  0.3$\rm \pm0.2^{0}$     \\ 
              & -2.81$\rm \pm0.04^{L22}$ &                   & -0.27$\rm \pm0.22^{L22}$& 0.37$\rm \pm0.06^{L22}$ & 0.24$\rm \pm0.08^{L22}$ & -0.16$\rm \pm0.16^{L22}$&  0.89$\rm \pm0.17^{L22}$ \\ \noalign{\smallskip}
KK            & -2.15$\rm \pm0.13^{0}$    & 0.15$\rm \pm0.02^{0}$& -0.05$\rm \pm0.12^{0}$& 0.6$\rm \pm0.06^{0}$      & 0.15$\rm \pm0.06^{0}$   & 0.35$\rm \pm0.08^{0}$   & 0.1$\rm \pm0.3^{0}$         \\
              & -1.84$\rm \pm0.05^{F20}$ &                   &                     &                     &                     &                     &               \\ \noalign{\smallskip}
\hline                                    
\end{tabular}
	\label{tab:3}
\end{table*}

\subsection{Calculating the Synthetic IL Spectra of GCs}
\label{sec:3_1} 

Synthetic IL spectra are computed with the \textit{CLUSTER} software package (\citetalias{Sh20}) in accordance with the selected abundances of chemical elements, stellar mass function, and stellar parameters T$_\text{eff}$ and log(g), defined by the chosen isochrone of stellar evolution. We use the plane-parallel hydrostatic models of stellar atmospheres set by the \textit{ATLAS~9} model grid \citep{Castelli03} based on the solar metal distribution from \citet{Grevesse98}\footnote{Lists of atomic and molecular spectral lines are available on the R. L. Kurucz website (\url{http://kurucz.harvard.edu/linelists.html}).}. We apply the air wavelengths for our analysis\footnote{The IAU standard for conversion from air to vacuum wavelengths is given in \citet{Morton91}.}. The program \textit{CLUSTER} has been developed with the methods of synthetic stellar spectra calculation described by \citet{Shimansky03}. This is performed in the course of numerical modelling of the radiative transfer in stellar atmospheres obtained by interpolation to the required parameters with the method described by \citet{Suleimanov96}. More specifically, cubic interpolation in T$_\text{eff}$ and linear in log(g) is used, which makes it possible to obtain the distributions of depth and gas pressure, as well as the spectrum of escaping radiation with an accuracy of 1\% \citep{Suleimanov96}. In the plane-parallel models of stellar atmospheres the distribution of temperature, pressure, and concentration of atoms and ions depending on the optical depth is determined by the given parameters of the star: T$_\text{eff}$, log(g), and metallicity. When computing the synthetic spectra of individual stars, for the given specific stellar parameters and chemical composition, the concentrations of atoms and ions are calculated depending on the depth in the atmosphere, and the contribution of each atom or ion to the emission and absorption coefficients at the given wavelength is determined. 
 
Model IL spectra of GCs ($I(\lambda$)) are computed using the synthetic spectra of the stars ($S(\lambda$,m)) according to the given mass function $\phi(m)$ with the initial mass (m) taken from the isochrones of stellar evolution: $$ \label{form:1} I(\lambda) = \int_{m1}^{m2} S(\lambda,m)\phi(m)dm.$$ We calculate synthetic IL spectra with the stellar mass function by \citet{Chabrier05} (formula 2). It is in agreement with both the theoretical distribution for low-mass stars \citep{Padoan04} and the \textit{HST}-observed mass function for nearby stars \citep{Zheng01}. In the mass range of stars in old globular clusters, this function is close to those derived by \citet{Salpeter55} and \citet{Kroupa01}. 

\subsection{Isochrone Selection for Calculating the IL Spectrum}
\label{sec:IsoSel}
\begin{table}
\centering
\caption{Chemical composition of the Sun in terms of the mass fractions of hydrogen (X), helium (Y), and metals (Z) adopted in our analysis with the \textit{ATLAS~9} model atmospheres \citep{Castelli03} and in stellar evolutionary models. See text for details.}
\begin{tabular}{llllc} \hline
Source           & X    & Y     & Z     &   $\rm Z/X$ \\ \hline \hline 
\textit{CLUSTER} with & 0.710 & 0.270 & 0.016 & 0.0230  \\ 
\textit{ATLAS~9} &       &       &       &         \\
Grevesse \&      & 0.716 & 0.266 & 0.018 & 0.0245 \\
Noels (1993):    &       &       &       &         \\ 
Grevesse \&      & 0.735 & 0.248 & 0.017 & 0.0230  \\
Sauval (1998)    &       &       &       &         \\  
Caffau et al.    & 0.732 & 0.253 & 0.0153 & 0.0209 \\ 
 (2011)          &       &       &       &         \\  
\hline \hline                                    
\end{tabular}
\label{tab:sol}
\end{table}
In this study, we mainly use the isochrones of stellar evolution by \citet{B08} (hereafter, \defcitealias{B08}{B08}\citetalias{B08})\footnote{\url{http://cdsweb.u-strasbg.fr/cgi-bin/qcat?J/A+A/484/815}} including the helium burning phase. The minimum initial stellar mass for the isochrones by \citetalias{B08} is $0.15~M_{\sun}$. The selected isochrones are based on the solar abundances from \citet{Grevesse93} and on the initial metallicity of the Sun $\rm Z = 0.017$ \citep{Grevesse98}, i.e. close to the abundance scale on which the \textit{ATLAS~9} code is based. Tab.~\ref{tab:sol} shows the mass fractions of hydrogen (X), helium (Y), and metals (Z) adopted in our analysis with the \textit{ATLAS~9} model atmospheres \citep{Castelli03}  compared to several compilations of the solar chemical composition \citep{Grevesse93, Grevesse98, Caffau}.
 The BASTI \citep{P04}\footnote{\url{http://albione.oa-teramo.inaf.it}} isochrones are based on the solar abundances from \citet{Grevesse93}. The new BASTI models \citep{H18, P21}\footnote{\url{http://basti-iac.oa-abruzzo.inaf.it/isocs.html}} employ the \citet{Caffau} solar metal distribution. One can ensure that the solar chemical element distribution adopted for our analysis is closer to the distributions by \citet{Grevesse93} and \citet{Grevesse98} than to the distribution by \citet{Caffau} used in the new BASTI models.

The temperature and luminosity of HB and other stars depend on many factors including He, CNO and metal abundances. Therefore, the change in the abundance scale in our analysis is not straightforward and will be studied elsewhere.
The \citetalias{B08} isochrones were computed with the mass loss by stellar wind during the red giant branch (RGB) of low-mass stars taken into account using the mass loss rate parameter \citet{Reimers75} $ \eta=0.35$. This value is close the $ \eta$ parameter range directly estimated with Kepler for HB stars in two old open clusters by \citet{Miglio12}. \citet{Miglio12} compared the difference between the average mass of low-luminosity RGB and red clump stars to theoretical predictions using, among others, \citetalias{B08} isochrones.

Selection of the isochrone for calculating the synthetic IL spectrum can be performed using an automatic procedure that minimizes the deviations between the synthetic spectrum and the observed one normalised to the synthetic. The program is written in the Python language v3.8.8\footnote{\url{http//www.python.org/}} using the Numpy and Scipy\footnote{\url{https://scipy.org}} packages. The description of the program is given in the Appendix~\ref{app:2} to the paper. For this purpose, a grid of the synthetic IL spectra is used calculated with the specified chemical composition, as well as the specified set of parameters of isochrones and [Fe/H] of the utilized  models of stellar atmospheres. 
 In the process of the isochrone assortment (see Appendix~\ref{app:2} for more details), 
the synthetic spectrum that best matches the observed one is selected from the pre-calculated grid ones by minimization of the $\rm \chi^2$ function:
\begin{equation} \rm \chi = \sum_{i=0}^{N}\left({\frac{obj_{i}-theor_{i}[q_1,q_2,q_3,q_4]}{err_{i}}}\right)^2, \end{equation}
where $\rm obj_i$ and $\rm err_i$ are the elements of the cluster observed spectrum and error spectrum, $\rm theor_i$ is a synthetic spectrum set by the isochrone parameters q1, q2, q3, and q4 (Y, the logarithm of age, the metallicity of the isochrone Z, and the metallicity of model atmospheres $\rm [Fe/H]_{atm}$). Before calculating $\chi$, the continuum level of the observed spectrum is normalised to the level of the synthetic continuum.

 To model the spectrum of the GC with the lowest metallicity in our sample, Ext~8 (Tab.~\ref{tab:2}), we used a scaled-solar canonical BASTI \citep{P04} isochrone with the mass loss rate parameter \citep{Reimers75} $ \eta=0.4$. In the BASTI canonical models by \citet{P04} based on the solar abundances from \citet{Grevesse93}, Y is related to metallicity as follows: dY/dZ $\sim$ 1.4. The BASTI isochrones with various values of Y at a given metallicity \citet{P06} are based on alpha-enhanced heavy elements distribution and do not include HBs. The alpha-enhanced canonical BASTI isochrones \citet{P06} with the opacities from \citet{Ferguson05} include HBs and are available only for the He-normal composition: dY/dZ $\sim$ 1.4. For the above mentioned reasons, in this study we use only the isochrones in the solar abundance scale by \citet{Grevesse93} or \citet{Grevesse98} which are close to each other. We do not use the new  BASTI isochrones \citep{H18, P21} based on the solar abundances from \citet{Caffau}.

To illustrate in more detail the reasons for our choice of isochrones, in Tab.~\ref{tab:iso_comp} (Appendix~\ref{app:4}) we compare $\rm T_{eff}$ and $\rm L (L_{\odot})$ for the three main stages of stellar evolution: the maximum in $\rm T_{eff}$ along the Main sequence (MSTO point), the tip of the red giant branch (TRGB), and the start of quiescent core He burning (HB). Isochrones with an age of 12.5~Gyr and $\rm Z=0.0001$ by \citetalias{B08}, by \citet{P04} and \citet{P06} (BASTI), and by \citet{H18} and \citet{P21} (BASTI~new) are considered. Pairwise compared isochrones differ only in one of the following parameters: the helium mass fraction (Y), the mass-loss efficiency $\eta$ according to the \citet{Reimers75} formula, and availability of taking into account the effect of atomic diffusion ($\rm Diff=Y$ or N).
We also compare the scaled-solar canonical BASTI isochrones by \citet{P04} and \citet{H18} with the equal parameters Y, $\eta$, and $\rm Diff$; 
the scaled-solar with the alpha-enhanced  canonical BASTI isochrones (\citet{P04} and \citet{P06}) with other similar parameters; and
the scaled-solar with the alpha-enhanced  new BASTI isochrones  (\citet{H18} and \citet{P21}) with other similar parameters.
Here are some conclusions that follow from this comparison.
When moving from the scaled-solar canonical BASTI isochrones \citep{P04} to the new scaled-solar BASTI isochrones \citep{H18} with all similar parameters ($\rm  Z=0.0001, Age=12.5~Gyr, Diff=N, \eta=0.3, Y\sim0.25$), significant changes are observed in $\rm T_{eff}$ in the MSTO and HB points and in $\rm L (L_{\odot})$ in the TRGB point (case (a) in Tab.~\ref{tab:iso_comp}). Namely, the new BASTI isochrones have a hotter MSTO by about 100~K and a cooler HB by about 160~K, as well as a lower TRGB luminosity by about $\rm 200 L_{\odot}$.  The authors notice \citep{H18}: "The main reason for the differences between these new BaSTI computations and the previous ones is the updated solar metal distribution and associated lower Z at a given $\rm [Fe/H]$."\footnote{ The lower luminosity of the core He-burning phase at old ages (1-12~Gyr) is caused by the use of the updated electron conduction opacities in the new BASTI models \citep{H18}.}

These significant variations in $\rm T_{eff}$ and $\rm L (L_{\odot})$ arising, when the initial solar chemical composition changes, impose restrictions on our use of isochrones. We will come back to other comparison results given in Appendix~\ref{app:4} later. 

\subsection{On the contribution of isochrone points to the IL spectrum of a GC}
\label{sec:3_3} 
 The grids of model atmospheres by \citet{Castelli03} that we use set the following parameter limitations for any [Fe/H] values: 0.0 $<$ log(g) $<$ 5.0 and temperature 3400K $<$ T$_\text{eff}$ $<$ 45000~K. If for the current isochrone point, one parameter goes beyond the specified limits, the nearest boundary value is assigned to it. This procedure does not affect the accuracy of calculations, since, as a rule, cool low-mass MS stars making a small contribution to the spectrum of the cluster go beyond the indicated ranges.

We found out that when analysing the low-resolution IL spectra of GCs it is acceptable to take into account only the isochrone points with a contribution larger than 0.2\% to the IL spectrum. The detailed description of the procedure for selecting the isochrone points is presented in Appendix~\ref{app:3_3}. Note that when using this procedure we do not change the original parameters of the isochrone points ($\rm T_{eff}$, $\rm log g$, $\rm R$, and $\rm M$). An example of the result selecting the isochrone points to calculate the IL spectrum is presented in Fig.~\ref{fig:S}, which argues that, when the number of points is reduced, the structure of the isochrone and the total relative contribution of the isochrone points to the overall spectrum are preserved. 

\begin{figure}
    \centering
	  \includegraphics[width=0.9\columnwidth]{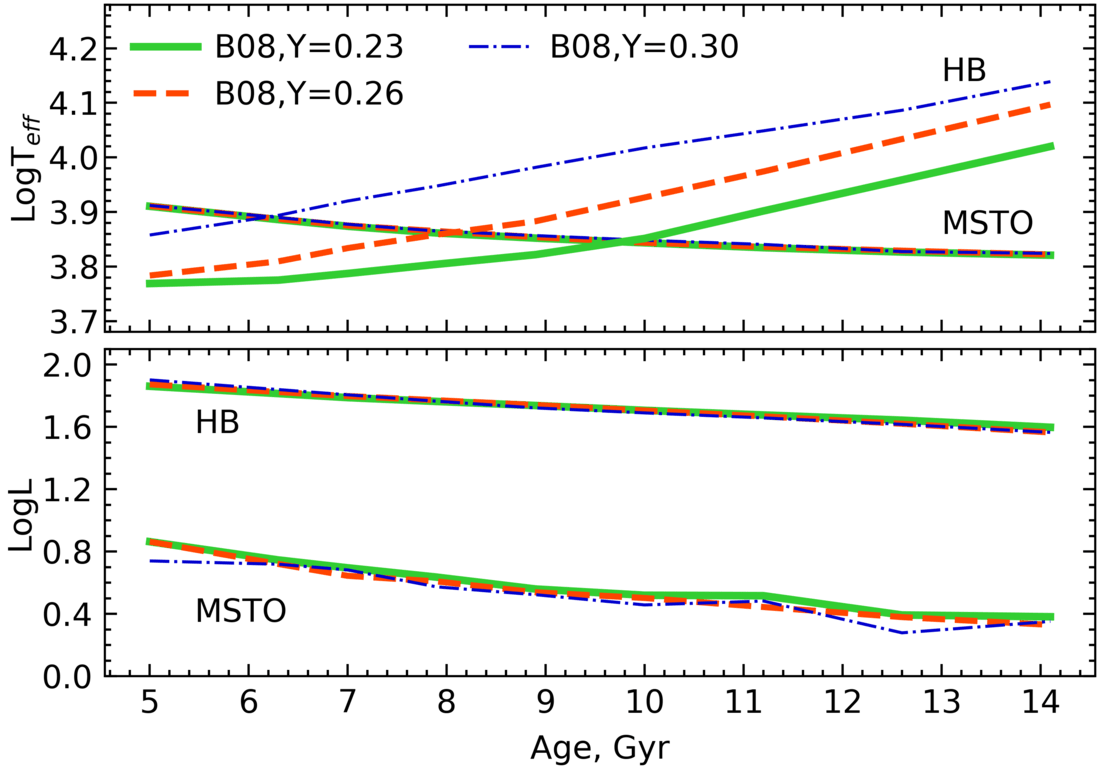}
    \caption{Effective temperatures and luminosities of the MSTO stars and the hottest HB stars depending on age. The \citetalias{B08} isochrones are used for Z=0.0001. The data for different isochrones are highlighted in colour as explained in the legend.}
    \label{fig:3}
\end{figure}
\begin{figure*}
    \centering
	  \includegraphics[width=0.9\textwidth]{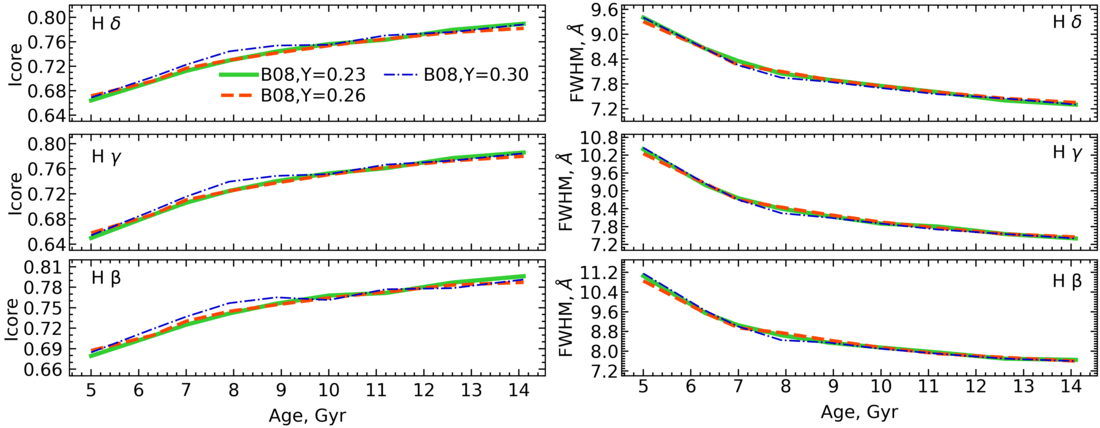} 
    \caption{Variation of I$_\text{core}$ and FWHM with age for three Balmer hydrogen lines in the synthetic IL spectra of GCs with the metallicity $\text{Z} = 0.0004$. The spectra were obtained using the \citetalias{B08} isochrones (the solid, dashed, and dash-dotted lines for $\text{Y} = 0.23$, 0.26, and 0.30, respectively) ignoring the HB stars.}
    \label{fig:4}
\end{figure*}

\begin{figure*}
    \centering
	  \includegraphics[width=0.9\textwidth]{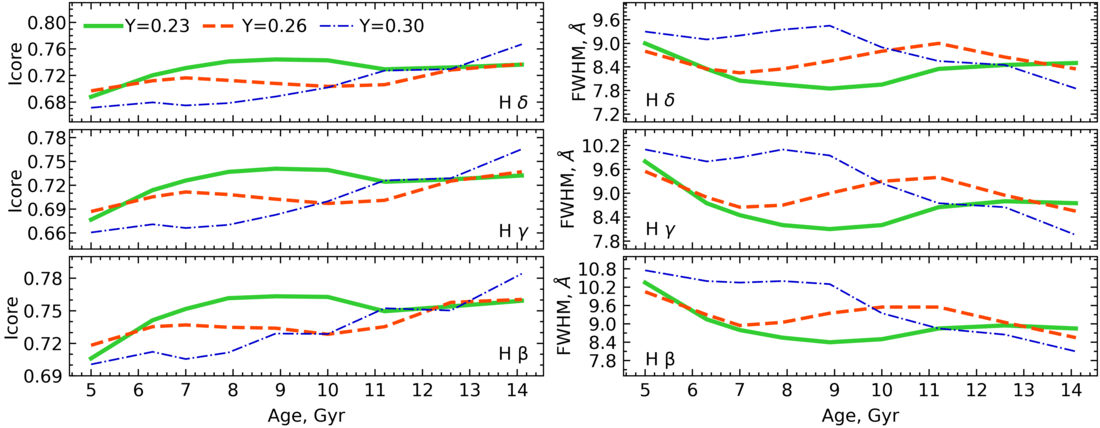}  
    \caption{Same as in the previous figure but with the HB stars included. }
    \label{fig:5}
\end{figure*}
 

\section{On the influence of the HB stars on the Balmer lines in the IL spectra}
\label{sec:4}
It has been established in a number of studies (see Sec.~\ref{Intro}), that HB stars make a significant contribution to the IL spectra of GCs.   

In the following, we will study the changes in the intensities, i.e. FWHM and depth (I$_\text{core}$), of the Balmer lines in the IL spectra of very low-metallicity GCs depending on the age. 
 Our experience shows that separate consideration of these two parameters characterizing absorption lines is useful due to the fact that stars at different evolutionary stages contribute in different proportions to the IL spectra, and FWHM and I$_\text{core}$ in the spectra of stars depend on the parameters of their atmospheres. For example, the hydrogen lines of hot HB stars are wide and deep, while the lines of RGB stars are narrow and shallow. FWHM and I$_\text{core}$ of the Main sequence and subgiant branch stars have different, intermediate values between those for the HB and RGB stars. In this section, we consider the isochrones with varying Y and $\text{Z} = 0.0001$ ($\text{[Fe/H]} \sim -2.23$~dex) (\citetalias{B08}). In the Appendix~\ref{app:1} we examine the isochrones by \citetalias{B08} with $\text{Z} = 0.0004$ ($\text{[Fe/H]} = -1.63$~dex).
Since, as a result of our analysis of the spectrum of EXT8, the metallicity of this GC turned out to be lower than the metallicities in the model grid of \citetalias{B08}, in this case we used the isochrone by \citet{P04} (left panel of Fig.\ref{fig:2}). 
Note that we measure the parameters FWHM and I$_\text{core}$ in the H$_{\delta}$, H$_{\gamma}$, and H$_{\beta}$ hydrogen lines normalised to unity in the wavelength regions: $4089.05 - 4115.4$~\AA, $4318.4 - 4363.5$~\AA, and $4815.8 - 4896.5$~\AA, respectively.

Figs.~\ref{fig:2}, \ref{fig:A2} in the Appendix~\ref{app:1} show how the isochrones vary with age and Y. In particular, with the increase of age, the effective temperatures of the MSTO stars decrease, while for the hottest HB stars, on the contrary, they increase. With the increase of age, the luminosity of the MSTO stars and of the hottest HB stars decrease. Similar conclusions concerning the variations with age are valid for the scale-Solar canonical BASTI isochrones \citet{P04} with $\text{Z} = 0.00001$ (left panel of Fig.~\ref{fig:2}). 

Fig.~\ref{fig:3} compares the variation of the parameters logT$_\text{eff}$ (on the left) and logL (on the right) with age for the MSTO stars and stars at the HB blue end (see also Fig.~\ref{fig:A3} in the Appendix~\ref{app:1} for the case of $\text{Z} = 0.0004$). The temperatures of the MSTO stars decrease with age. The luminosities of the MSTO stars and the hottest HB stars behave similarly. On the other hand, logT$_\text{eff}$ for the stars at the HB hot end increases non-linearly with age at different rates for different Y. If we consider the \citetalias{B08} isochrone, then the higher the Y value, the hotter the HB first point.

The considered variations of the parameters logT$_\text{eff}$ and logL significantly affect the FWHM and I$_\text{core}$ of the Balmer lines in the spectra of GCs which will be illustrated below. 
Fig.~\ref{fig:4} shows the variation of I$_\text{core}$ and FWHM with age for three Balmer hydrogen lines in the synthetic IL spectra of GCs with the metallicity $\text{Z} = 0.0001$ (see also Fig.~\ref{fig:A4} in the Appendix~\ref{app:1} for the case of $\text{Z} = 0.0004$).  The model dependences in Figs~\ref{fig:4} and \ref{fig:5} are built for the spectral resolution $\text{FWHM} = 5.5$~\AA. The synthetic spectra were obtained using the \citetalias{B08} isochrones for $\text{Y} = 0.23$, 0.26, and 0.30 (Fig.~\ref{fig:5}). Fig.~\ref{fig:4} shows that FWHM and I$_\text{core}$ vary monotonously with age. The higher the value of I$_\text{core}$, the shallower the line becomes, i.e., closer to the continuum level. 
Similar changes of  I$_\text{core}$ and FWHM with age for three Balmer hydrogen lines can be seen in the case of $\text{Z} = 0.0004$ (Fig.~\ref{fig:A4} in the Appendix~\ref{app:1}). 

If helium burning stages of stellar evolution are taken into account when building the IL synthetic spectra (Fig.~\ref{fig:5}), the behaviour of the FWHM and I$_\text{core}$ variation is no longer monotonous with age. 
First, the depth and FWHM decrease with the increase of age, mainly due to the fact that T$_\text{eff}$ and L of the MSTO stars decrease, but T$_\text{eff}$ of the HB stars does not increase in such a quick way as to compensate for this decrease.
\begin{figure*}
    \centering
	 \includegraphics[width=0.45\textwidth]{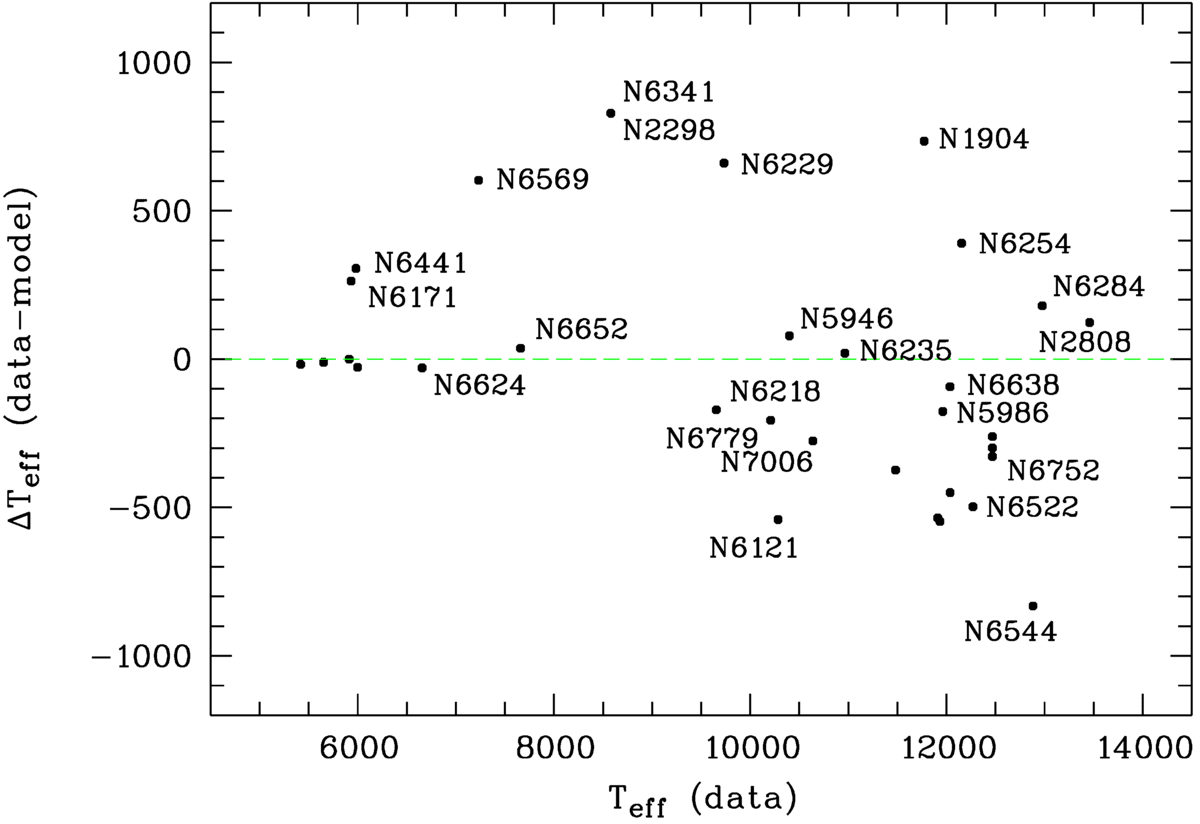} 
      \includegraphics[width=0.45\textwidth]{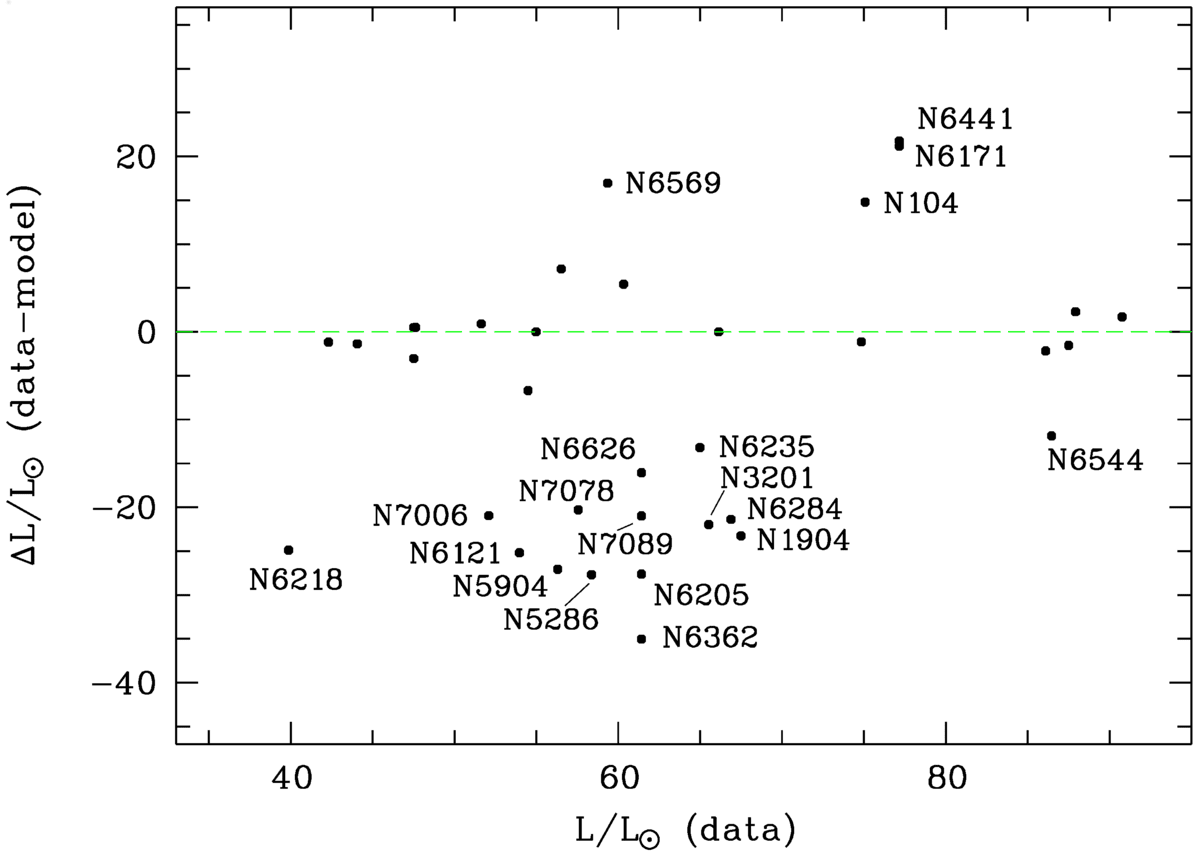} 
    \caption{Differences between $\rm T_{eff}$ (left) and $\rm log L (L_{\sun})$ (right) observed for the cluster HB stars and the isochronous ones \citetalias{B08}  (see Sec.~\ref{sec:5_1} for details). The isochrones, with which the observed CMDs were compared, were selected by \citetalias{Sh20} examining the IL spectra of the GCs from the library of \citet{Sch05}. The observed $\rm T_{eff}$ and $\rm log L (L_{\sun})$ are plotted along the X axes in the diagrams.
}
    \label{fig:T_L}
\end{figure*}

As soon as $\rm T_{eff}$ for the hottest HB stars exceed T$_\text{eff}$ for the MSTO stars, and as T$_\text{eff}$ of the hottest HB stars continues to increase (see Fig.~\ref{fig:2} in the Appendix~\ref{app:1} and Fig.~\ref{fig:3}), the hydrogen lines start getting deeper and wider. This continues until the moment, when T$_\text{eff}$ reaches $\sim$ 9000~K for the HB stars, at which the hydrogen line intensities begin to decrease again. This is due to the intense hydrogen ionization at the given temperature. 
Thus, the growth rate of FWHM and I$_\text{core}$ for the HB stars varies depending on age. This is due to how quickly their temperature and contribution to the IL spectrum varies with age. 
Similar changes of  I$_\text{core}$ and FWHM with age for three Balmer hydrogen lines can be seen in the case of using the isochrones by \citetalias{B08} with HBs and $\text{Z} = 0.0004$ (Fig.~\ref{fig:A5} in the Appendix~\ref{app:1}).

 \citet{Lee00} and \citet{Percival11} detected a non-monotonic variation of the Lick $\rm H_{\beta}$ index, when hot HB stars were considered. The details of the modelling of the horizontal branch morphologies are different in our and those studies. The general conclusion is that underestimation of the contribution of the HB stars to the IL spectra results in underestimation of the age of GCs. In the work by \citet{Percival11}, there is an important conclusion about the difficulty of taking into account statistical fluctuations in the number of stars on the HB in the spectra of real GCs when determining their age and metallicity. \citet{Percival11} found that the extension of the real HB in the CMD, and accordingly, the $\rm H_{\beta}$ indices differ from the modelled ones by setting an average mass and implementing the Gaussian spread in masses of individual stars coming on to the HB for each stellar evolutionary model.    

\section{Results and Discussion}
\label{sec:5}  
\subsection{Comparison of the Distribution of Stars on the Observed CMDs with the Isochrones Selected when Modelling the IL Spectra}
\label{sec:5_1} 
\begin{figure}
    \centering
	  \includegraphics[width=0.8\columnwidth]{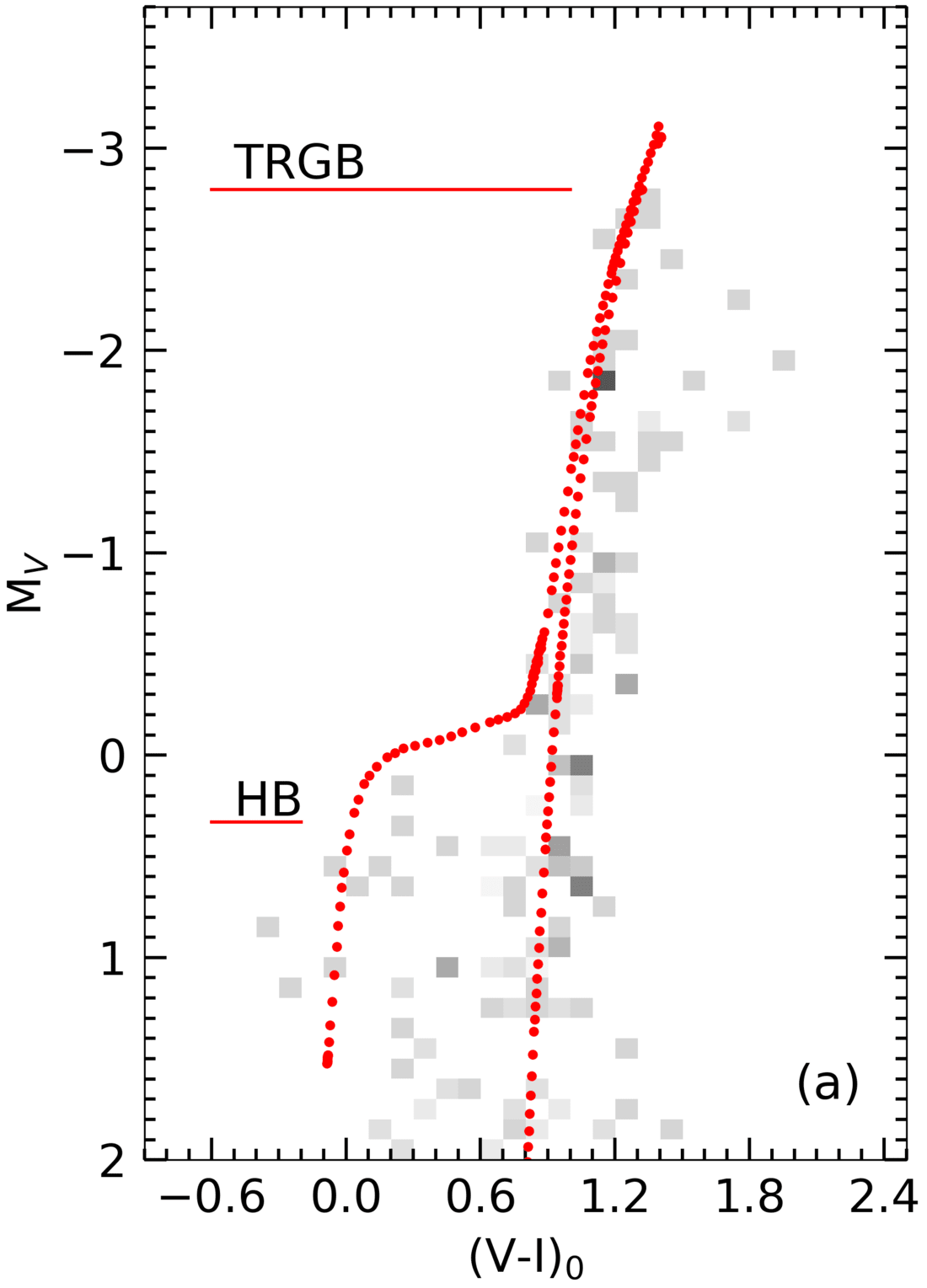} 
         \includegraphics[width=0.8\columnwidth]{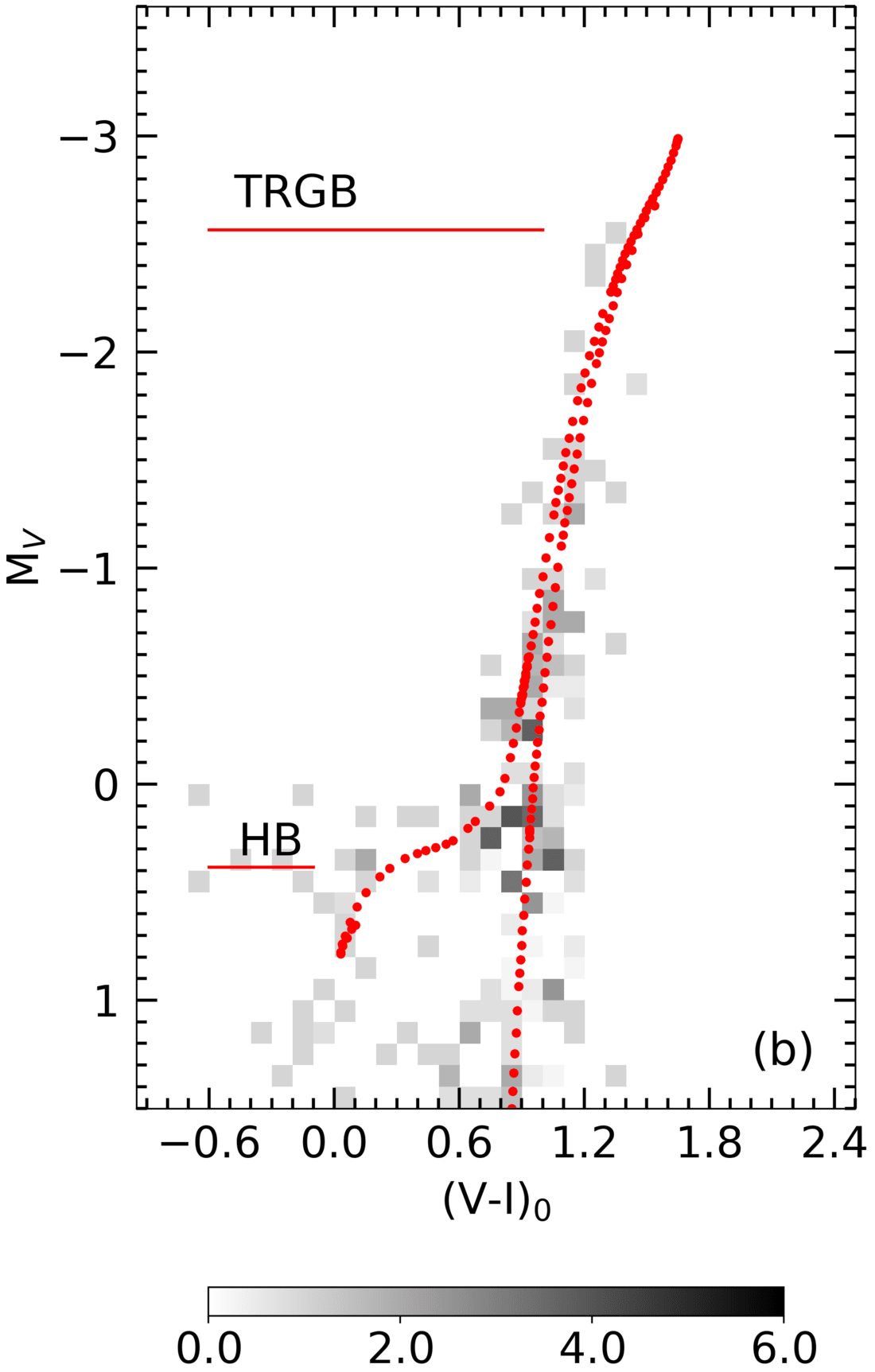} 
    \caption{Comparison of the isochrones selected in the IL spectrum modelling (Table~\ref{tab:2}) with the Hess diagrams built from the observed CMDs of the clusters H~III (a) and B~317 (b) (see Sec.~\ref{sec:5_1_1} for details).}
    \label{fig:7}
\end{figure}
\begin{table*}
	\centering
	\caption{Obtained by us (marked with superscript $^{0}$) and from the literature: 
	(2) the estimated apparent magnitude of the HB level,
	(3) the absolute magnitude of the HB level calculated with formula (2) from \citet{F12} and our [Fe/H] estimates from Table~\ref{tab:2},
	(4) the estimated colour excess according to \citet{Schlafly11}, and (5) the distance modulus.  
	The literature data are from the papers by: \citet{L21}~(L21), \citet{Fan08}~(F08), and \citet{McConnachie05}~(M05).}
 \begin{tabular}{lcccr}       
\hline                                                                
Parameter/   &  $\rm V^0$          &   $\rm M_V^0$ & $\rm E(B-V)$           & $\rm (m-M)_0$              \\                         
 Object      &  (HB)               &    (HB)       &                        &                            \\                                     
 \hline                                                                      
HIII         &  25.25$\rm \pm0.19$ &  0.33         & 0.17$\rm \pm0.15^{0}$  & 24.36$\rm \pm 0.24^{0}$   \\                              
             &                     &               & 0.175$\rm ^{M05}$      & 24.15$\rm ^{M05}$            \\  
B317         &  25.25$\rm \pm0.17$ &   0.385       & 0.08$\rm \pm0.13^{0}$ & 24.64$\rm \pm 0.2^{0}$        \\                                 
             &                     &               &  0.11$\rm ^{F08}$      &                              \\ 
EXT8         &  24.89              &  0.19         &  0.06$\rm ^{L21}$      & 24.43$\div$24.57$\rm ^{L21}$ \\                     
\hline 
\end{tabular}
\label{tab:4}
\end{table*}

 Before considering the comparison of isochrones with the parameters determined from analysing the IL spectra and the stellar photometry data of extragalactic GCs available in the literature, let us turn to such a comparison for the Galactic GCs analysed by \citetalias{Sh20}. Accurate stellar photometry for them was performed by \citet{Piotto02} and by \citet{Sar07} in the B, V and V, I filters of the Johnson Cousins photometric system, respectively. Following \citetalias{Sh20}, we plot CMDs of the GCs as colour corrected for extinction versus absolute magnitude, i.e., $\rm (B-V)_0$ versus $M_V$ \citep{Piotto02}, or $(V-I)_0$ versus $\rm M_I$ \citep{Sar07}.  Then we superimpose the \citetalias{B08} isochrones with the parameters determined by \citetalias{Sh20} onto the CMDs and estimate for 35 Galactic GCs the differences between $\rm T_{eff}$ and $\rm L (L_{\sun})$ observed for the cluster HB stars and the isochronous ones.
 We estimate $\rm T_{eff}$ of bright hot stars on the blue end of the HB and $\rm log L (L_{\sun})$ of the mid-HB stars by interpolation of theoretical relations for HB stars: colour versus $\rm T_{eff}$ and absolute magnitude versus $\rm log L (L_{\sun})$. A detailed description of this procedure is given in the Appendix~\ref{app:3}. While this is a crude approximation, taking into account, among other factors, the scatter of the colours and magnitudes of HB stars, this comparison with deep photometry data will help us to reveal the caveats and limitations of our method.

Figure \ref{fig:T_L} shows the estimated and observed $\rm T_{eff}$ and $\rm L (L_{\sun})$ values in comparison with the corresponding theoretical ones. The standard deviation from the isochronous luminosity value in the I filter for the mid-HB stars has a scatter of about $\rm std (I_{HB}) = 0.19\pm0.1$~mag.  
On average over the sample of 35 Galactic GCs, this value corresponds to: $\rm \Delta L \sim 15 L_{\sun}$ (Fig.~\ref{fig:T_L}). It should be noted that for half of the sample objects, this value does not exceed $\Delta L \sim 4 L_{\sun}$. 
 When considering the results of the comparison of the isochrones in Tab.~\ref{tab:iso_comp} (Appendix~\ref{app:4}), discussed in Sec.~\ref{sec:IsoSel},
the largest discrepancy in $\rm L (L_{\sun})$ for the HB is expected between the isochrones with different $\eta$. The isochrones with the higher $\eta$ demonstrate the lower $\rm log L (L_{\sun})$ for HB (case (f) in Tab.~\ref{tab:iso_comp}). Therefore, one of the reasons for the differences in $\rm log L (L_{\sun})$ in the selected isochrones and the data for the real HB stars on the CMD may be the difference in the average values $\eta$ for a cluster from those isochronous.
Another reason for large deviations of the luminosity of HB stars from the corresponding isochronous values ($\rm \Delta L \ge 10 L_{\sun}$) may be the contribution of hot stars of GCs, beyond the HB luminosity, such as blue straggler stars, asymptotic giant branch (AGB), and variable stars to the IL spectrum. Also, field stars of the corresponding temperature that do not belong to GCs can contribute.

The average colour scatter ($(V-I)_0$, $\rm (B-V)_0$) of bright hot stars on the blue end of the HB for 35 Galactic GCs is $\sim$0.1~mag. The typical deviation of the average colours ($\rm T_{eff}$) of stars in the HB blue part from the corresponding values at the isochrone points selected from the IL spectra of clusters by \citetalias{Sh20} is smaller than one percent for the HBs cooler than $\sim$8000~K, and~3\% for HBs with $\rm T_{eff}\sim10000$~K (Fig.~\ref{fig:T_L}). For hotter HBs, the deviation can reach 4-6\%. However, there are GCs with HBs cooler than $\sim$10000 K and with large $\rm \Delta T_{eff}$ (Fig.~\ref{fig:T_L}). These are GCs with complex HB morphology, such as NGC6171, NGC6441, and NGC6569. These GCs exhibit a rich-populated red part of the HB and a small number of stars in the blue part of the HB, which may nevertheless fall into the IL spectrum. 
 It follows from Tab.~\ref{tab:iso_comp} (Appendix~\ref{app:4}) that the greatest differences in $\rm T_{eff}$ of the HB blue end, up to 2000~K, occur between the isochrones with different Y and the isochrones with different $\eta$ and other equal parameters (cases (c) - (f) in Tab.~\ref{tab:iso_comp}). Therefore, the reason for the large deviations of the isochronous $\rm T_{eff}$ of the HB blue end from the average $\rm T_{eff}$ estimated for the blue end HB stars may be differences in the average values of $\eta$ and Y for a cluster from those isochronous.
Significant deviations from the model values of $\rm T_{eff}$ and $\rm log L (L_{\sun})$ occur for clusters close to the Galactic plane, as well as those shielded from us by the Galactic plane. Probability of the contribution of field stars to such IL spectra increases significantly. The parameters of isochrones selected from the IL spectra and, therefore, $\rm T_{eff}$ and $\rm log L (L_{\sun})$ will turn out to be distorted for the main evolutionary stages, if field stars of any temperature and luminosity have contributed to the IL spectrum. In general, based on the sample of 35 Galactic GCs, we can conclude that for $\sim$75\% of the sample it is possible to determine the parameters of HB stars quite confidently with regard to the above-mentioned caveats. Similar problems can be expected for extragalactic GCs, especially for those observed in dense fields of field stars. 

It should be noted that, in addition to contamination problems, limitations on the accuracy of selecting the stellar evolution isochrone to describe the observed IL spectrum can be imposed by the specific features of the stellar evolution models used. For example, inclusion of atomic diffusion in the BASTI models \citep{P21, H18} significantly improved the result of approximating theoretical isochrones to the results of photometry of the MSTO stars. On the other hand, atomic diffusion increases the mass of the He-core at the He ignition for a given initial chemical composition and decreases the He abundance in the stellar envelope. The luminosity and $\rm T_{eff}$ of core He-burning stars are higher in \citet{P21} than those in the previous $\alpha$-enhanced BASTI models \citep{P06}. In Tab.~\ref{tab:iso_comp} (case (g)) (Appendix~\ref{app:4}) we compare the BASTI new \citep{H18} scaled-solar isochrones with taking or not into account the effect of atomic diffusion ($\rm Diff=Y$ or N). It can be seen that $\rm T_{eff}$ of the HB blue end is higher in the case of $\rm Diff=Y$ by approximately 400~K from that in the case $\rm Diff=N$.
 Another limitation is that it is not yet possible to take into account variations of some parameters during stellar evolution, for example, the variety of the convective overshoot conditions in stars treated in the models of the stellar evolution according to the mixing length theory by \citet{BV58} \citep[see, e.g.,][and references therein]{Viani18, Da18}. The mixing length and the mass-loss efficiency parameters are assigned fixed values in the models. However, in real GCs they can vary from star to star. Recently, the first direct estimates of the mass loss of the Galactic GCs on the RGB using asteroseismology methods have appeared. \citep{Howell23}. 
 
 Using Fig. \ref{fig:3}, as well as Figs. \ref{fig:2}, \ref{fig:A2}, and \ref{fig:A3} in the Appendix \ref{app:1}, one can estimate the maximum $\rm T_{eff}$ of HB stars belonging to our sample GCs in accordance with the isochrones selected from the IL spectra (Table~\ref{tab:2}). The most blue extended HBs host GCs HIII, C39, and Ext~8 with the maximum $\rm T_{eff}\sim12600K$. The nuclear GC in KK197 hosts hot HBs stars with $\rm T_{eff}\sim10000K$. The maximum $\rm T_{eff}$ of other four GCs (PA, B317, B2, and B165) is 8000-9000~K.


\subsubsection{HIII and B317}
\label{sec:5_1_1}
In the following we will check the correspondence of the isochrones that we selected, when modelling the IL spectra of the GCs, to the observed CMDs of these objects.
 We built the Hess diagrams for HIII and B317 (Fig.~\ref{fig:7}) using the stellar photometry by \citet{Sh06}\footnote{The identifiers of Hodge~II and III were swapped around by mistake by \citet{Sh06}. Hence, the data for Hodge~II in their paper refer to Hodge~III and vice versa.} performed with the images from WFPC2 HST. The circular apertures for selecting stars within and around the clusters remained the same as those determined in the original paper. The colour intensity in each bin (0.1~mag $\times$ 0.1~mag) on the Hess diagrams means the number of stars labelled under the colour-bars at the bottom of the diagram. Fig.~\ref{fig:7} demonstrates the results of subtracting the diagram for the field stars from the diagram for all the stars in the cluster region. The corresponding \citetalias{B08} isochrones for HIII and B317 from Table~\ref{tab:2} are overplotted.
 One can see that reasonable agreement exists between the data and the models. The panels are provided with the isochrone parameters, as well as the distance modulus $\rm (m-M)_0$, and the colour excess $\rm E(B-V)$\footnote{The extinction in a particular bandpass is described by $\rm A_{\lambda}=R_V \cdot E(B-V)$, where $\rm E(B-V)$ is colour excess and $\rm R_V$ is equal 3.1 for the Milky Way type extinction law \citep{Schlafly11}.} \citep{Schlafly11} used to match the isochrones to the observed distribution of stars on the CMD. To determine $\rm (m-M)_0$ and $\rm E(B-V)$, the following characteristics were considered: the average colour of the RGB at $\sim$ 0.5~mag above the HB and the HB luminosity level calculated with formula (2) from \citet{F12} (see Table~\ref{tab:4}, column~4). These benchmarks are quite acceptable for solving the problem, since the number of stars near the TRGB of the clusters is small and the field stars, making a significant contribution in terms of population to the CMD, differ little in colour and luminosity from the stars belonging to the objects under study. Table~\ref{tab:4} shows our and literature $\rm (m-M)_0$ and $\rm E(B-V)$, as well as the absolute magnitude of the HB level (column~5) calculated with formula (2) from \citet{F12} and our [Fe/H] estimates from Table~\ref{tab:2}. The second column gives the corresponding apparent magnitude of the HB level (a thin solid line in Fig.~\ref{fig:7} ) calculated from M$_\text{V}$(HB), $\rm (m-M)_0$, and $\rm E(B-V)$ in Table~\ref{tab:4}.

\subsubsection{EXT8}
\label{sec:5_1_2}
\begin{figure}
   \centering
       \includegraphics[width=0.46\textwidth]{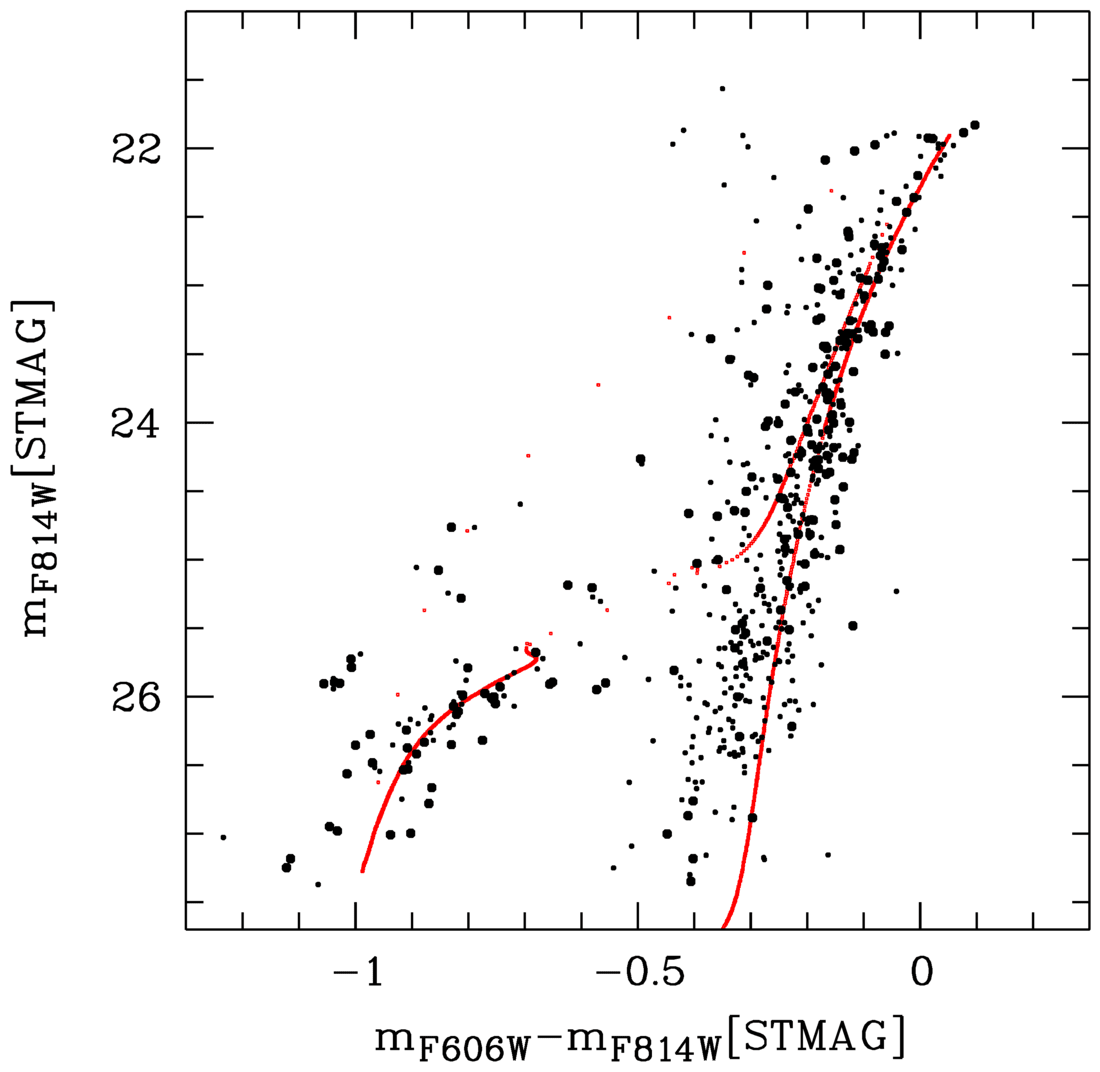} 
    \caption{Comparison of the isochrone selected in the modelling of the EXT8 spectrum (Table~\ref{tab:2}) shown in red and the photometry results for this cluster from \citet{L21} (see Section~\ref{sec:5_1_2} for details).}
    \label{fig:8}
\end{figure}
 
EXT8 is remote from M31 GC residing in the halo of this galaxy. Its deep CMD \citet{L21} (hereafter: \defcitealias{L21}{L21}\citetalias{L21}) 
is much less polluted by field stars and the photometric depth is greater than in the case of HIII and B317. There are quite a lot of stars near the TRGB on the CMD of EXT8, and the HB is clearly visible (\citetalias{L21}).
Fig.~\ref{fig:8} shows the CMD of EXT8 built from the \citetalias{L21} results in a radius of 30~pc (200~pix $\times$ 0.15~pc/pixel) from the centre of the object. The large black dots in Fig.~\ref{fig:8} demonstrate the stellar photometry in the drc images in the F606W and F814W filters\footnote{table a2 from cdsarc.u-strasbg.fr (130.79.128.5)} selected by photometric errors (dF606W $<$ 0.1~mag and dF814W $<$ 0.1~mag) and by the ALLFRAME sharpness parameter which was set to be in the range from $-0.1$ to 0.22. After this selection, the ALLFRAME chi-square parameter for the stellar images in two filters appears to be smaller than 13. Note that the photometry table in the flc images from \citetalias{L21}\footnote{table a1 from 130.79.128.5} does not contain the sharpness and chi-square parameters. Therefore, the selection for these data was performed only by the photometry errors. The small black dots in Fig.~\ref{fig:8} demonstrate the stellar photometric data in the flc images from \citetalias{L21} selected by the photometric errors as follows. We have set the errors on the magnitudes in two $HST$ filters (dF606W and dF814W) and R.M.S. on the F606W and F814W magnitudes to be smaller than 0.1~mag. Such a limitation on the photometry errors was not chosen by chance, since according to artificial star experiments in \citetalias{L21} (see their fig.~3), the photometric errors should be smaller than 0.1~mag for the RGB and HB stars. 
In addition to the RGB and HB stars, the stars bluer than the RGB as well as the stars brighter than the HB, which are apparently the field stars, remain in the cluster region after the selections. In the mF814W filter, the stars being in terms of brightness approximately at the level of the TRGB and bluer than it are located mainly in the central part of EXT8 and have, on average, high values of chi-square and sharpness.
\begin{figure*}
    \centering
	  \includegraphics[width=1.0\textwidth]{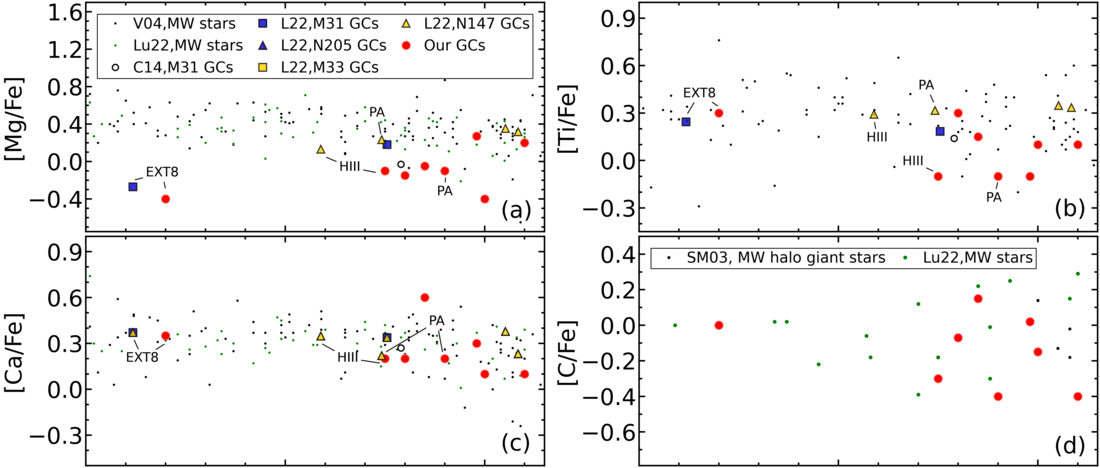} 
         \includegraphics[width=1.0\textwidth]{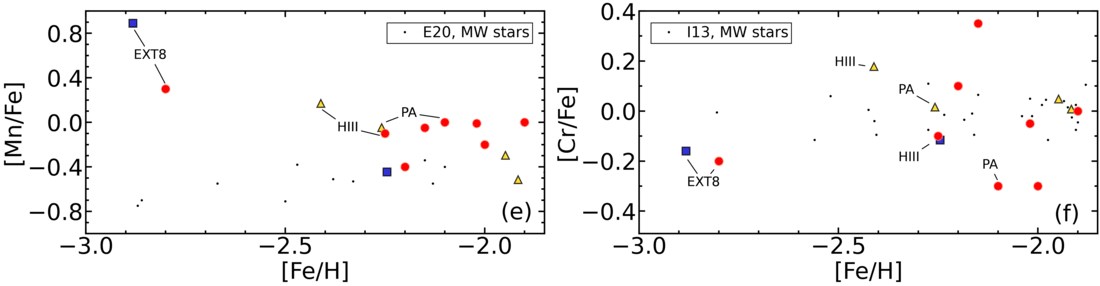} 
    \caption{LTE abundances of chemical elements for the stars in our Galaxy (the small dots) and for GCs in M31 and in its neighbours (the large symbols) obtained from analysing their IL spectra. The values obtained in this paper are shown by the red circles. The open circles and triangles mark the abundances for the M31 cluster from \citet{S16} (S16) and \citet{Col14} (C14), respectively. The coloured squares show the LTE abundances for the M31 cluster and its vicinity from \citet{L22} (L22). The small dots indicate the data for the Galactic field stars from \citet{Lu22} (Lu22) (panels~(a),(c),(d)), from \citet{Venn04} (V04) (panels~(a),(b),(c)), from \citet{Smith03} (SM03) (panel~(d)), from \citet{E20} (E20) (panel~(e)), and from \citet{Ishigaki13} (I13) (panel~(f)).}
    \label{fig:10}
\end{figure*}

It can be seen in Fig.~\ref{fig:8} that the solar-scaled canonical BASTI isochrone \citep{P04} chosen by us for analysing the EXT8 IL spectrum with the parameters $\text{Z} = 0.00001$, $\text{Age} = 11$~Gyr, and $\text{Y} = 0.245$ adequately describes the RGB and HB clusters. It should be noted though that the metallicity of this isochrone $\text{[Fe/H]} \sim -3.23$~dex is lower than the value obtained from the analysis of the IL spectra $\text{[Fe/H]} \sim -2.8$~dex (Table~\ref{tab:3}). 
We used $(\text{m} - \text{M})_0 = 24.5$~mag and $\rm E(B-V)$ $=$ 0.06~mag obtained by \citetalias{L21}. To convert the BASTI isochrone data from VEGAMAG WFC3 UVIS to the STMAG system, we used the corrections from \citetalias{L21}: $\Delta_{F606W} = 0.246$~mag and $\Delta_{F814W} = 1.259$~mag. 
Note that the instability strip in the selected isochrone (Fig.~\ref{fig:8}) approximately coincides with that on the CMD and with that found by \citetalias{L21}. The colour boundaries for the instability strip: M$_{F606W}$ $-$ M$_{F814W}$ from $-0.8$ to $-0.5$ (\citetalias{L21}), or m$_{F606W}$ $-$ m$_{F814W}$ from $-0.736$ to $-0.436$, since A$_{F606W} = 0.168$~mag and A$_{F814W} = 0.104$~mag.  
This analysis suggests that our choice of the isochrone (Table~\ref{tab:2}) is not bad. Let us turn to Tab.~\ref{tab:iso_comp}(case (j)) (Appendix~\ref{app:4}). It demonstrates a comparison of $\rm T_{eff}$ and $\rm L (L_{\sun})$ for the MSTO, TRGB, and HB in the scaled-solar canonical BASTI isochrones \citep{P04} with the metallicities $[Fe/H]=-3.27$~dex and $[Fe/H]=-2.27$~dex and other equal parameters used by us in the analysis of the IL spectrum of Ext~8. The value $[Fe/H]=-2.8$ dex is approximately halfway between these two metallicities.  If we divide the differences in $\rm T_{eff}$ and $\rm L (L_{\odot})$ by 2, then we can notice the following most significant differences between the isochrones. The HB is cooler by $\sim 65$~K and brighter by $\rm \sim 60 L_{\odot}$, and the luminosity of the TRGB is higher by about $\rm \sim100 L_{\odot}$ for the isochrone with $[Fe/H]=-2.8$~dex compared to the isochrone used. Since the agreement between the observed and model spectra in the region of the Balmer lines is not ideal (Fig.~\ref{fig:1}), especially in the core and wings of the $\rm H_{\beta}$ line, then these discrepancies can be attributed to the above-mentioned differences in $\rm T_{eff}$ and $\rm L (L_{\sun})$ for the MSTO, TRGB, and HB.  

\subsection{Chemical Composition of the sample GCs}
\label{sec:5_2}

Fig.~\ref{fig:10} shows a comparison of the chemical composition for the sample GCs with the corresponding abundances obtained with high-resolution spectroscopy for the field stars of the Galaxy \citep{Lu22, E20, Ishigaki13, Smith03, Venn04} and GCs of M31 from the papers by \citet{L22}, \citet{Col14}, and \citet{S16}. 
An overall agreement between the abundances for our sample GCs and the literature data can be seen. However, it should be noted that there are differences $\leq0.2$~dex between the literature and our LTE abundances for GCs (see also \ref{tab:3}). The exception is the lower [Mn/Fe] value we obtained compared to that by \citet{L22} (\ref{fig:10}, panel~e). We estimated the Mn abundance using the blend near 4033\AA\ (see Fig. 1). The total contribution of the intense Mn I 4030.76, 4031.79, 4033.07, 4033.58, 4033.65, 4034.49, and 4035.72 \AA\ lines to this blend generally about 1.5 times exceeds the contribution of Fe. It should be noted that, in general, [Mn/Fe] for GCs is on average higher than that for field stars. Low Mn abundances in dwarf galaxies served as indication of sub-Chandrasekhar-mass white dwarf progenitors of SNe Ia in these objects \citep{Reyes20}.

It should also be noted that almost all GCs of our sample, except B317 and B165, have low ($\rm [Mg/Fe]\lesssim 0$) in comparison with field stars for which $\rm [Mg/Fe]\sim 0.4$~dex.  Mg is the so-called $\alpha$-element mainly produced by explosive C-burning in massive stars and in core-collapse supernovae (CCSNe). Unlike Mg, Ca, whose abundance appears normal compared to field stars (Fig.~\ref{fig:10}), is mainly produced in CCSNe and SNe~Ia. Low [Mg/Fe] values for GCs obtained from IL spectra were considered as evidence for the presence of multiple stellar populations in GCs \citep[e.g.][]{L22}. Low $\rm [Mg/Fe]\sim -0.24$~dex were obtained by \citet{Sh18} from the IL spectra of low-metallicity Galactic GCs NGC~6341 and NGC~7078 observed with the CARELEC spectrograph at the 1.93-m telescope of the Haute-Provence Observatory. These results are consistent with the discovery by \citet{Masseron19} of extreme Mg-depleted stars in these GCs and of Mg-Al anticorrelation among the clusters' stars. The depletion reaches $\rm [Mg/Fe]\sim-0.5$~dex. Even stronger Mg-depletion likely exists for Ext~8 and B2, as follows from the analysis of their IL spectra (\citet{L22}, Table~\ref{tab:3} this paper).  

$\rm [C/Fe]$ values that we determined from the IL spectra in the optical range, are consistent with that of the field stars, but on average higher than $\rm [C/Fe]$ for RGB stars in Galactic GCs \citep{Roediger14} and for GCs in M31 from IL spectroscopic observations in the H band \citep{S16}. The main contribution to the IL spectrum in this wavelength range is made by RGB stars. On the other hand, IL spectra in the optical range include the contribution from all the cluster stars. Note that MSTO stars make a significant contribution to the IL spectrum in the optical range.
It is known from the literature that the abundances of light elements: Li, C, N, and O in stellar atmospheres change during the dredge-up processes on the RGB \citep{Kraft94, Gratton00}. Thus, the difference between $\rm [C/Fe]$ values obtained from analysing IL spectra of GCs and the corresponding C abundances of RGB stars in GCs are due to the change of the chemical composition of the stellar atmospheres during their evolution.  

Regarding the accuracy of [C/Fe] estimates using low-resolution IL spectra, it should be noted that the contribution of carbon to the CH band at $4240-4330$~\AA\ is approximately five times greater than the contribution of oxygen. The abundances of other elements (Ti, Mg, Si, Ca, Al, and Fe), whose lines contribute the CH band, can be determined independently using other absorption features in the spectrum. An additional test for the correctness of [C/Fe] obtained from the CH band is provided by other molecular lines involving C, for example, CN band at $4120-4220$~\AA. There are no O-dominated spectral features in the studied spectra. Therefore, [O/Fe] is estimated indirectly and depends on [C/Fe]. The abundance of oxygen was set within the limits: [O/Fe] $\sim$ 0.3 $\div$ 0.5~dex. In \citetalias{Sh20}, it was demonstrated that [C/Fe] for the Galactic GCs obtained with our method is higher by about 0.4~dex than those obtained with high-resolution spectroscopy for some of the brightest stars in these clusters. However, there are no systematic differences between IL [C/Fe] abundances obtained using the same spectra of Galactic GCs from \citet{Sch05} by \citet{Conroy18} and \citetalias{Sh20}. 

\section{Concluding remarks}
\label{Summary}

 The distribution of GCs by colour and metallicity is bimodal in many massive galaxies with prominent metallicity peaks near $\text{[Fe/H]} = -1.6$~dex and $\text{[Fe/H]} = -0.6$~dex \citep{Harris10, Beasley19}. Very low metallicity GCs are rare, massive, and predominantly have blue HBs \citep{Harris10, Beasley19}. Their chemical composition and chemical anomalies among their stars is a subject of extensive studies. Recently, extremely low-metallicity GC Ext~8 in M31 has been discovered \citep{L22}. This discovery changed our understanding about the so-called 'metallicity floor', i.e., the minimum metallicity of a GC surviving in the Universe \citep{Beasley19}. Our study contributes to investigation of very metal-poor GCs. We determine the properties of HBs, i.e., $\rm T_{eff}$ and luminosity of HB stars set by the isochrones of stellar evolution for GCs with $\rm [Fe/H]\le -2$~dex. We improve the method by \citetalias{Sh20} developed for determination of the age, metallicity Z, specific helium abundance Y, and the abundances of several chemical elements: Fe, C, Mg, Ca, Mn, Ti, and Cr. Using the low-resolution IL spectra of GCs, we explore how the FWHM and the depth of the Balmer absorption lines change depending on the isochrone used and on the properties of HB stars. We show a comparison between the parameters of HB stars on the isochrones used to model the IL spectra (\citetalias{Sh20}) and the corresponding observed characteristics of HB stars from the Galactic GCs (see details in Sec.~\ref{sec:5_1}). This comparison allows us to identify caveats of the employed method. We conclude that the main sources of errors in the isochrone selection with our method of modelling the intensities of the Balmer hydrogen lines in the IL spectra of GCs are the lack of knowledge about the actual mass losses of the RGB stars in the clusters under study (we used the isochrones with the fixed mass-loss efficiency parameters $ \eta$ \citep{Reimers75}), foreground and background contamination and stochastic variations in the number of stars at different evolutionary stages within the studied aperture. 

  Our research has shown that all GCs in the sample are old ($10 \leq T \leq 13.6$~Gyr) and have blue HBs. Using the dependences between  $\rm T_{eff}$ and the age (Fig.~\ref{fig:3} and Figs.~\ref{fig:2}, \ref{fig:A2} and \ref{fig:A3} in the Appendix~\ref{app:1}), we estimated the maximum $\rm T_{eff}$ of HB stars belonging to our sample GCs, in accordance with the isochrones selected from the IL spectra (Table~\ref{tab:2}). Studying the IL spectra of HIII, C39, and Ext~8 reveals that they have blue extended HBs with the maximum $\rm T_{eff}\sim12600K$.
 The nuclear GC in KK197 hosts hot HBs stars with $\rm T_{eff}\sim10000K$. The maximum $\rm T_{eff}$ of HB stars in other four GCs (PA, B317, B2, and B165) is 8000-9000~K.

Chemical abundances of Ca, Ti, C, and Cr correspond well to the abundances of stars in the Galactic field (Sec.~\ref{sec:5_2} and Fig.~\ref{fig:10}). Carbon abundances determined from the IL spectra of GCs are higher than the corresponding data for RGB stars in the Galactic GCs \citep{Roediger14} and for GCs in M31 from the IL spectroscopic observations in the H band \citep{S16}.
This is due to the change in chemical composition of stellar atmospheres during their evolution (Sec.~\ref{sec:5_2}). Almost all GCs of our sample, except for B317 and B165, have low Mg abundances ($\rm [Mg/Fe]\lesssim 0$) compared to field stars, whose $\rm [Mg/Fe]\sim 0.4$~dex. This is an indication of the presence of multiple populations \citep{L22} and of extremely Mg-depleted stars \citep{Masseron19} in these GCs. Manganese abundances are higher in our sample GCs, than those in field stars. Low Mn abundances in dwarf galaxies were explained by the presence of sub-Chandrasekhar-mass white dwarf progenitors of SNe Ia \citep{Reyes20}.


It can be concluded that there are similarities between the properties of very low-metallicity ancient GCs in different galaxies and different environments, which indicates similar evolutionary processes. However, individual parameters of stars in GCs, such as $\rm T_{eff}$, luminosity, and chemical composition, undergo changes depending on local physical conditions. 

\section*{Acknowledgements}
We thank the anonymous referee for comments that helped to improve the paper.
Spectroscopic observations of the object KK197-2 reported in this paper were obtained with the SALT under programmes 2020-1-RSA~OTH-005 and 2019-1-RSA~OTH-003 (PI: Kniazev). 
This work is partly supported by the grant of the Ministry of Science and Higher Education of the Russian Federation no. 075-15-2022-262 (13.MNPMU.21.0003). 

\section*{Data Availability}
 Data used and shown in this article are available on reasonable request to the corresponding author. 





\clearpage
\appendix

\clearpage

\section{Description of the program for selecting the theoretical isochrone for the optimum description of the observed IL spectrum of the cluster}
\label{app:2}

To implement the algorithm, a grid of the synthetic spectra was calculated using the method described in Section~\ref{sec:3}. The resolution of the model spectra is $ \rm FWHM= 0.00166$~\AA. 
The model spectra are smoothed to the resolution of the observed spectrum (FWHM $=$ 5.5~\AA\ in our case). The estimates of the logarithm of the \citetalias{B08} isochrone age for calculating the grid of the model spectra were chosen as follows: from 9.7 to 10.15 in 0.05 increments. The Y values were: 0.23, 0.26, and 0.30. For the IL spectrum of the cluster obtained from telescope observations and reduced as described in Section~\ref{sec:2}, the error spectrum is calculated using the formula: $\rm \sqrt{(obj+sky) Gain\cdot npix+Ron^2 npix}$, where obj+sky is the one-dimensional spectrum of the object before the sky subtraction, obtained in the same aperture as the analysed cluster spectrum and converted into the wavelength scale using the same dispersion relation; Gain is the gain factor of the CCD in [e-/ADU]; npix is the FWHM of the cluster in observations; Ron is the CCD readout noise in electrons. The spectrum of the object after reducing and subtracting the sky is also converted into electrons by multiplying by Gain.

At the first stage of the algorithm, a synthetic spectrum is searched for from the pre-calculated grid ones with a minimum deviation from the observed one, according to the parameter:

\[
\rm \chi = \sum_{i=0}^{N}\left({\frac{obj_{i}-theor_{i}[q_1,q_2,q_3,q_4]}{err_{i}}}\right)^2,
\]

where $\rm obj_i$ and $\rm err_i$ are the elements of the cluster observed spectrum and error spectrum, $\rm theor_i$ is a synthetic spectrum set by the isochrone parameters q1, q2, q3, and q4 (Y, the logarithm of age, the metallicity of the isochrone Z$_{B08}$, and the metallicity of model atmospheres $\rm [Fe/H]_{atm}$). Before calculating $\chi$, the continuum level of the observed spectrum is normalised to the level of the synthetic continuum. The parameters of the found synthetic spectrum are used as an initial approximation at the next stage of the program. At the next stage, a non-linear least-squares problem is solved with the given limitations of the parameters. To do this, the built-in function of the  \textit{scipy} library was used: \url{scipy.optimize.least_squares}\footnote{ \url{ https://docs.scipy.org/doc/scipy/reference/generated/scipy.optimize.least_squares.html}} which finds the local minimum of the function F(x): F(x) $=$ 0.5 $\cdot$ $\rm \Sigma_{i=0}^{m-1} (\rho(f_i(x)^2))$, under the condition of $ \rm lb \leq x \leq ub$, where f(x) is the residual function $\rm f_i(x) = (obj_i - theor_i(x))/err_i$, $\rm i$ is the spectrum element; and $\rm \rho$(s) is the loss function. This function uses the default value: $\rm \rho(s) = s = f(x)^2$, where x are the required parameters: Y, log(Age), Z$_{B08}$, and [Fe/H]$_{atm}$. Minimization is performed using the `trf' (Trust Region Reflective) algorithm. At each iteration of the minimization algorithm, the model spectrum (theor(x)) is calculated as a linear interpolation of the grid spectra to the obtained parameters x. The observed cluster spectrum (obj) and the error spectrum (err) are normalised to the model interpolated spectrum (theor(x)) also at each iteration. The program calculates the 95\% confidence interval for the obtained parameters. The procedure for normalising the spectra to the theoretical continuum level is as follows. A list of wavelengths, in which the continuum is determined, was compiled in advance. The intensities of the observed spectrum at these points of the continuum are averaged over the range of $\rm \pm$1~\AA. To determine the continuum level of the model spectrum, the maximum intensity in the range of $\rm \pm$1~\AA\ is taken. The resulting continuum points are linearly interpolated over the whole length of the observed and model spectra. As a result, the observed spectrum is divided by the continuum of the observed spectrum and multiplied by the pseudo-continuum of the model spectrum.

\begin{figure*}
    \centering
      \includegraphics[width=1.0\textwidth]{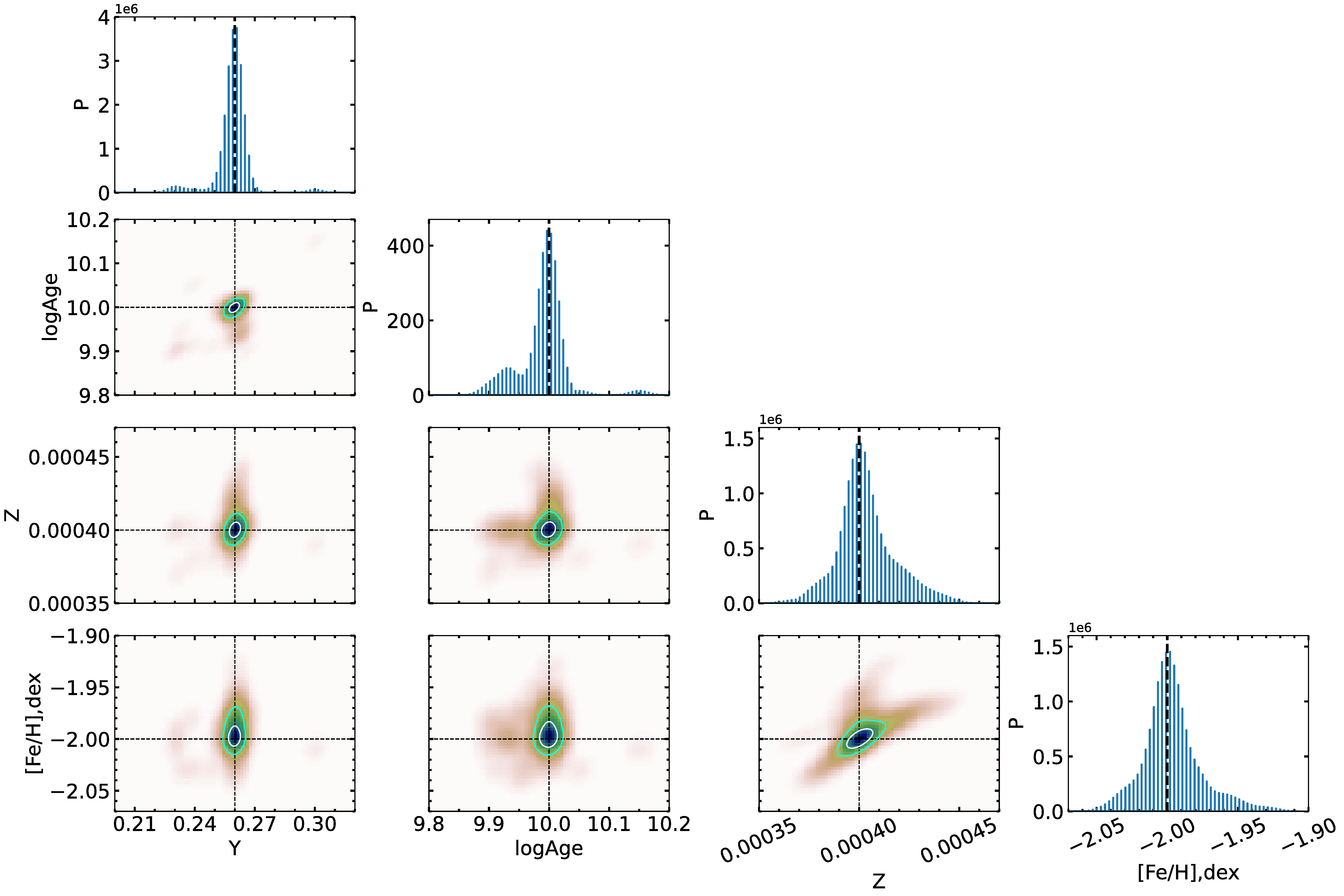}
    \caption{Results of the Monte Carlo experiments (see Appendix~\ref{app:2}) for details.}
    \label{fig:B1}
\end{figure*}

Fig.~\ref{fig:B1} shows the results of experiments on the selection of the isochrone parameters for a hundred of spectra obtained from one synthetic one calculated with the \citetalias{B08} isochrone parameters $\rm Y = 0.26$, $\rm log(Age) = 10.0$, $\rm Z = 0.0004$, and metallicity of model atmospheres $\rm [Fe/H] = -2.0$~dex by adding random noise using the Monte Carlo method. The average SNR in each spectrum is 100. In this case, when artificially noisy model spectra are used as the studied spectra, the error spectra are random noise which is on average 1/100 with respect to signal. 
Four diagonal panels in Fig.~\ref{fig:B1} show the probability distributions for each of the parameters: Y, log(Age), Z, and [Fe/H].
Other panels of Fig.~\ref{fig:B1} show the joint probability distributions for these parameters computed with kernel density estimates using the Gaussian kernels.
The 1$\sigma$ and 2$\sigma$ error contours are drawn in the panels of Fig.~\ref{fig:B1}. The dispersions of the parameters that we have determined are as follows:  $\sigma =$ 0.008, 0.034, 1.43E-5 and 0.018 for Y, log(Age), Z and [Fe/H], respectively. 
The cross marks the parameters of the isochrone of the synthetic spectrum used in the experiment. 
Fig.~\ref{fig:B1} shows the correlation between the parameters Y and log(Age) determined by the algorithm, as well as the correlation between the metallicity of the isochrone Z and the metallicity of the model atmospheres [Fe/H].  
The correlation between Z and [Fe/H] can be explained as follows. The mass fraction of chemical elements heavier than helium, Z, includes abundances of many chemical elements including Fe and iron peak elements\footnote{The Fe abundance in solar units: $\rm [Fe/H] $=$ log(N_{Fe}/N_H) - log(N_{Fe}/N_H)_{\sun}$, where $\rm N_{Fe}/N_H$ -- the ratio of the iron and hydrogen concentrations by the number of atoms, or by mass, is related to the mass fraction of chemical elements heavier than helium, Z. Obviously, the sum of mass fractions of hydrogen, helium, and metals: $\rm X + Y + Z = 1$.}. The correlation between Y and log(Age) indicates the difficulty of separating these parameters while analysing the IL spectra. This issue was discussed in Section~\ref{sec:4} in more detail.

The program and the set of synthetic spectra computed with the \citetalias{B08} isochrones and a fixed chemical composition will be accessible via GitHub.  

\newpage
\section{Comparison of stellar evolution models used in this study}
\label{app:4} 
 \begin{table*}
\centering
\caption{Results of comparison of $\rm T_{eff}$ and $\rm L (L_{\odot})$ for the three main evolutionary stages (MSTO, TRGB, and HB). The isochrones with an age of 12.5~Gyr and $\rm Z=0.0001$ by \citetalias{B08}, by \citet{P04} and \citet{P06} (BASTI), and by \citet{H18} and \citet{P21} (BASTI~new) are considered. 
 Subscripts SS and $\alpha$ mean that we consider the scaled-solar or alpha-enhanced isochrones. See the text for details.}
 \begin{tabular}{l|ccc|ccr} 
\hline \hline
Isochrone (1)                                         & \multicolumn{3}{c}{$\rm T_{eff}(1) - T_{eff}(2)$} &  \multicolumn{3}{c}{$\rm L(1) -  L(2)$} \\ 
isochrone (2)                                         &   MSTO   &  TRGB   &   HB                &     MSTO   &  TRGB   &   HB          \\  
\hline \hline                                        
   (a)                                                 &          &         &                     &            &         &              \\
(1) BASTI$\rm_{SS} (Y=0.245, \eta=0.3)$                & -110.05  &  -36.5  & 156.3               &  -0.13     & 210.8   & 1.67         \\
(2) BASTI~new$\rm_{SS} (Y=0.247, Diff=N, \eta=0.3)$    &          &         &                     &            &         &              \\ \noalign{\smallskip}
  (b)                                                  &          &         &                     &            &         &              \\
(1)  BASTI$\rm_{SS} (Y=0.245, \eta=0.4)$               &   31.14  & -32.6   &  1005.4             &   0.07     &  -10.1  &  -2.75       \\
(2) B08$\rm_{SS}(Y=0.26, \eta=0.35)$                   &          &         &                     &            &         &              \\ \noalign{\smallskip}
 (c)                                                   &          &         &                     &            &         &              \\ 
(1) B08$\rm_{SS}(Y=0.30, \eta=0.35)$                   &  45.2 &  16.3&  1727.9 &  -0.04 &  -74.5 &  -0.80  \\ 
(2) B08$\rm_{SS}(Y=0.26, \eta=0.35) $                  &          &         &                     &            &         &              \\ \noalign{\smallskip}
   (d)                                                 &          &         &                     &            &         &              \\
(1) B08$\rm_{SS}(Y=0.26, \eta=0.35)$                   &  38.7 &  19.2 &  1954.5 &  -0.07 &  -114.4 &  -3.02  \\ 
(2) B08$\rm_{SS}(Y=0.23, \eta=0.35) $                  &          &         &                     &            &         &              \\ \noalign{\smallskip}
 (e)                                                   &          &         &                     &            &         &              \\ 
(1) BASTI~new$\rm_{\alpha} (Y=0.30, Diff=N, \eta=0.3)$ & -30.5    & 24.1    &  2306.2             & -0.19      & -89.3   & -2.95        \\
(2) BASTI~new$\rm_{\alpha} (Y=0.247, Diff=N, \eta=0.3)$&          &         &                     &            &         &              \\ \noalign{\smallskip}
   (f)                                                 &          &         &                     &            &         &              \\
(1) BASTI~new$\rm_{SS} (Y=0.247, Diff=N, \eta=0.3)$    &  -9.7    &  -18.65 &   2171.1            &  -0.03     &  -10.77 &   -13.7      \\
(2) BASTI~new$\rm_{SS} (Y=0.247, Diff=N, \eta=0.0)$    &          &         &                     &            &         &              \\ \noalign{\smallskip}
   (g)                                                 &          &         &                     &            &         &              \\
(1) BASTI~new$\rm_{SS} (Y=0.247, Diff=N, \eta=0.3)$    & 291.7    &  10.7   &   -425.2            &   0.37     &  -19.12 &   2.96       \\
(2) BASTI~new$\rm_{SS} (Y=0.247, Diff=Y, \eta=0.3)$    &          &         &                     &            &         &              \\ \noalign{\smallskip}
 (h)                                                   &          &         &                     &            &         &              \\
(1) BASTI$\rm_{SS} (Y=0.245, \eta=0.4)$                &  -5.35   & -37.2   &  -56.4              &   0.02     & -3.3    & 0.07         \\
(2) BASTI$\rm_{\alpha} (Y=0.247, \eta=0.4)$            &          &         &                     &            &         &              \\ \noalign{\smallskip}
  (i)                                                  &          &         &                     &            &         &              \\
(1) BASTI~new$\rm_{SS} (Y=0.247, Diff=N, \eta=0.3)$    &   -8.07  & -9.8    & -23.5               &  -0.04     & 12.6    & -0.47        \\
(2) BASTI~new$\rm_{\alpha} (Y=0.247, Diff=N, \eta=0.3)$&          &         &                     &            &         &              \\ \noalign{\smallskip} 
\hline
(j)  &          &         &                     &            &         &              \\
(1) BASTI$\rm_{SS} (Z=0.0001, [Fe/H]=-2.27, Y=0.247, \eta=0.4)$             & -30.06 & -61.73 & 130.86  & -0.14 & 249.57 & 130.87   \\
(2) BASTI$\rm_{SS} (Z=0.00001, [Fe/H]=-3.27, Y=0.247,\eta=0.4)$             &        &        &         &       &        &          \\ \noalign{\smallskip}
\hline  \hline
\end{tabular}
\label{tab:iso_comp}
\end{table*}

In Tab.~\ref{tab:iso_comp}, $\rm T_{eff}$ and $\rm L (L_{\odot})$ are compared for the three main stages of stellar evolution: the maximum in Teff along the MS (the MSTO point), the TRGB, and the start of quiescent core He burning (HB). The isochrones are considered which have an age of 12.5~Gyr and $\rm Z=0.0001$ by \citetalias{B08}, by \citet{P04} and \citet{P06} (BASTI), and by \citet{H18} and \citet{P21} (BASTI~new). The remaining isochrone parameters under consideration are listed in parentheses in the first column of the table. Pairwise compared isochrones differ only in one of the following parameters: the helium mass fraction (Y), as in cases (c), (d), and (e); the mass-loss efficiency $\eta$ according to the \citet{Reimers75} formula, as in case (f); the flag of the treatment the effect of atomic diffusion ($\rm Diff=Y$ or N), as in case (g). In case (a), the scaled-solar canonical BASTI isochrones by \citet{P04} and \citet{H18} with equal parameters Y, $\eta$, and $\rm Diff$ are compared. In case (h), the scaled-solar and alpha-enhanced canonical BASTI isochrones (\citet{P04} and \citet{P06}) with other similar parameters are compared. In case (i), the scaled-solar and alpha-enhanced BASTI~new isochrones (\citet{H18} and \citet{P21}) with other similar parameters are compared. 

The following conclusions can be drawn when considering Tab.~\ref{tab:iso_comp}. 

(case a) When moving from the scaled-solar canonical BASTI isochrones \citep{P04} to the new scaled-solar BASTI isochrones \citep{H18} with all similar parameters ($\rm Z=0.0001, Age=12.5 Gyr, Diff=N, \eta=0.3, Y\sim0.25$), significant changes are observed in $\rm T_{eff}$ at the MSTO and HB points, and in $\rm L (L_{\odot})$ at the TRGB point. Namely, the new BASTI isochrones have a hotter MSTO by about 100 K and a cooler HB by about 160 K, as well as a lower TRGB luminosity by about $\rm 200 L_{\odot}$. As the authors notice, such variations are mainly a consequence of the change in the initial Solar chemical composition, on which the models are based\footnote{The lower luminosity of the core He-burning phase at old ages is caused by the use of the updated electron conduction opacities in the new BASTI models \citep{H18}.}. Since there are no isochrones with $\eta=0.3$ in the scaled-solar canonical BASTI \citep{P04} models, we averaged $\rm T_{eff}$ and $\rm L (L_{\odot})$ for the three main evolutionary stages, correspondingly, using the canonical isochrones with $\eta=0.2$ and $\eta=0.4$. 

(case b) Significant changes of $\rm T_{eff}$ at the HB take place when moving from the scaled-solar isochrones by \citetalias{B08} to the scaled-solar canonical BASTI isochrones \citep{P04}. With all similar parameters, including the initial Solar chemical composition, these models differ in their input physics, which, as can be seen, primarily affects $\rm T_{eff}$ for the HB. Pairwise comparison of the remaining corresponding isochrone parameters in case (b) does not show large differences. Since these models are based on the same initial Solar chemical composition, there may not be so large differences in $\rm T_{eff}$ at the HB when comparing other \citetalias{B08} and scaled-solar canonical BASTI \citep{P04} isochrones with all equal parameters. 

The largest differences in $\rm T_{eff}$ at the HB between the models are expected with the change of Y (cases c, d and e) or $\eta$ (case f). With the increase of Y, or the increase of the mass loss from $\eta=0$ to $\eta=0.3$, $\rm T_{eff}$ for the HB increases by about 2000 K. Additionally, as Y increases, $\rm L (L_{\odot})$ at the TRGB point decreases by approximately $\rm 100 L_{\odot}$, and $\rm T_{eff}$ also changes at the MSTO by approximately 30-50~K. With the increase of $\eta$, no other significant changes are observed besides the growth of $\rm T_{eff}$ at the HB point. 

When moving from the scaled-solar to alpha-enhanced isochrones by the same authors, no major changes in $\rm T_{eff}$ and $\rm L (L_{\odot})$ are observed for the three main evolutionary stages (cases h and i). 

(j) In this case, a comparison of the scaled-solar canonical BASTI isochrones \citep{P04} with different metallicities and other equal parameters is considered. This comparison was performed in order to approximately estimate the differences in $\rm T_{eff}$ and $\rm L (L_{\odot})$ for the three stages of stellar evolution during the transition from the BASTI isochrone \citep{P04} with $[Fe/H]=-3.2$ dex, which we used to model the spectrum of Ext~8 (Tab. \ref{tab:2}) to the BASTI isochrone \citep{P04} with $[Fe/H]=-2.8$~dex. We obtained the latter value when modeling the intensity of the Fe lines in the spectrum of Ext~8 (Tab. \ref{tab:3}). The metallicity $[Fe/H]=-2.8$ dex is approximately halfway between the successive (1 and 2) values of $[Fe/H]$ in Tab. \ref{tab:iso_comp} (case (j)). If we divide the differences in $\rm T_{eff}$ and $\rm L (L_{\odot})$ by 2, then we can notice the following most significant differences between the isochrones. The HB is cooler by $\sim 65$~K and brighter by $\rm \sim 60 L_{\odot}$, and the luminosity of the TRGB is higher by about $\rm \sim100 L_{\odot}$ for the isochrone with $[Fe/H]=-2.8$~dex compared to the isochrone used.

\clearpage
\section{Selection of Isochrone Points for Spectrum Modelling}
\label{app:3_3} 

\begin{figure}
    \centering
	  \includegraphics[width=0.9\columnwidth]{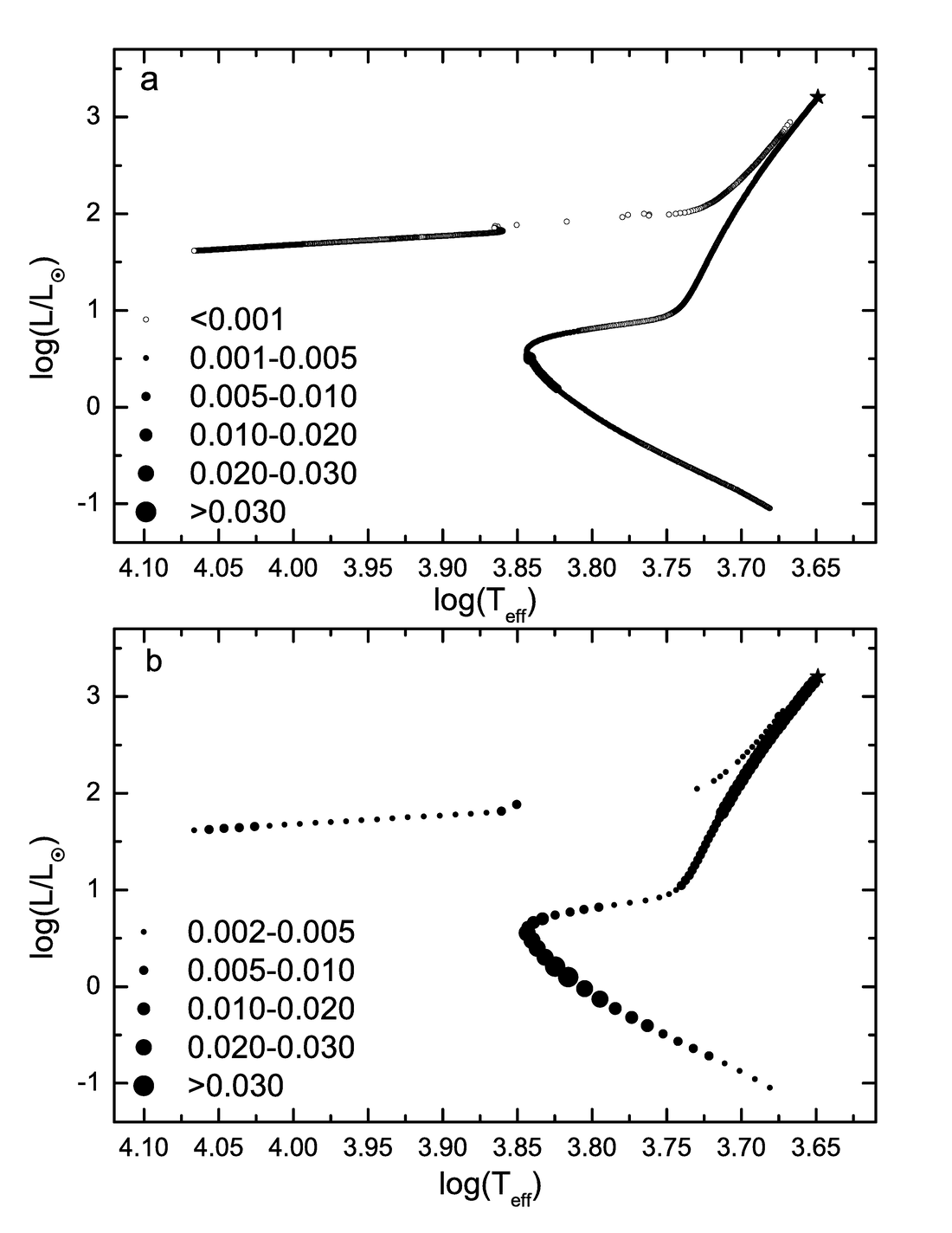}  
    \caption{Illustration of the result of selecting the isochrone points to calculate the IL spectrum (see Appendix~\ref{app:3_3} for details). The original distribution of the isochrone points according to their $T_{eff}$ and $\log g$ is shown in panel (a). The resulting distribution is shown in panel (b).}
    \label{fig:S}
\end{figure}

 When working with the \citetalias{B08} isochrones, we have developed the following procedure allowing us to check the significance of the contribution of each isochrone point to the IL spectrum of a GC and to possibly exclude the points with a very small contribution from our calculations. It should be noted that the masses at consecutive points often coincide in the isochrones by \citetalias{B08}. This is due to the fact that the mass changes slowly along the RGB and the AGB, and the values of the evolving masses are printed with not enough decimal digits.  As a result, the term $\rm dm$ in formula~\ref{form:1} is equal to zero.
   
The isochrone points with coinciding mass values near the TRGB, or on the AGB are of particular concern. Such stars have high luminosity. If a mass at the certain isochrone point coincide with the previous one, then the contribution of the next isochrone point to the IL spectrum may turn out to be implausibly large, greater than 5$\%$. In addition to solving the above issues, development of a procedure for selecting the isochrone points, when modelling IL spectra, helps reduce calculation time. 

The total luminosity of stars at the given wavelength $\lambda$ for an isochrone point with the number n can be approximately estimated as follows: 
\begin{equation}\label{form:2} \rm L(\lambda)_n = 4 \pi \sigma B(\lambda)_n (R_n)^2 F(M_n) \Delta M_n, 
\end{equation} 
where $\rm L(\lambda)$ is the luminosity to be found at the given wavelength, $\rm B(\lambda)_n$ is the Planck function, $\rm F(M_n)$ is the mass function, $\rm \Delta M_n = (M_{n+1} - M_n) / 2$ is half of the mass interval between consecutive points. In this study, we estimated the contribution of points to the IL of the cluster in the continuum at the wavelength $\rm \lambda = 5000$ \AA. The points with a contribution smaller than $0.2$ \% to the IL spectrum of a GC were excluded from consideration. The removed points should not differ in temperature by no more than $\rm \Delta \log T_{eff}= 0.01$ dex and in surface gravity by no more $\rm \Delta \log g= 0.06$ dex from neighbouring ones. These selection conditions were obtained experimentally. It should be noted that, when an isochrone point is removed, the contribution of the next point to the IL spectrum increases due to the increase of its $\rm \Delta M$ value. Therefore, two consecutive points could not be excluded simultaneously. The process of excluding the points was iterative, and before each iteration, the values $\rm L(\lambda)_n$ of all the points were calculated over again. Note, that in this process we do not change the original parameters of the isochrone points ($\rm T_{eff}$, $\rm log g$, $\rm R$, and $\rm M$). An example of the result of the described procedure is presented in Fig.~\ref{fig:S}. Panel (a) presents the initial distribution of the isochrone points according to their $\rm T_{eff}$ and $\rm \log g$. Panel (b) shows the points selected for calculating the synthetic IL spectrum. The circles of various sizes on the panel legends of Fig.~\ref{fig:S} denote the contribution of the points in the IL spectrum. The asterisk marks the RGB tip. Fig.~\ref{fig:S}~(b) argues that, when the number of points is reduced, the structure of the isochrone and the total relative contribution of the isochrone points to the overall spectrum are preserved.

\clearpage
\section{Additional material on the subject of the influence of the properties of HB stars on the intensities of the Balmer lines in IL spectra of GCs}
\label{app:1}
\begin{figure*}
    \centering
	  \includegraphics[width=0.8\textwidth]{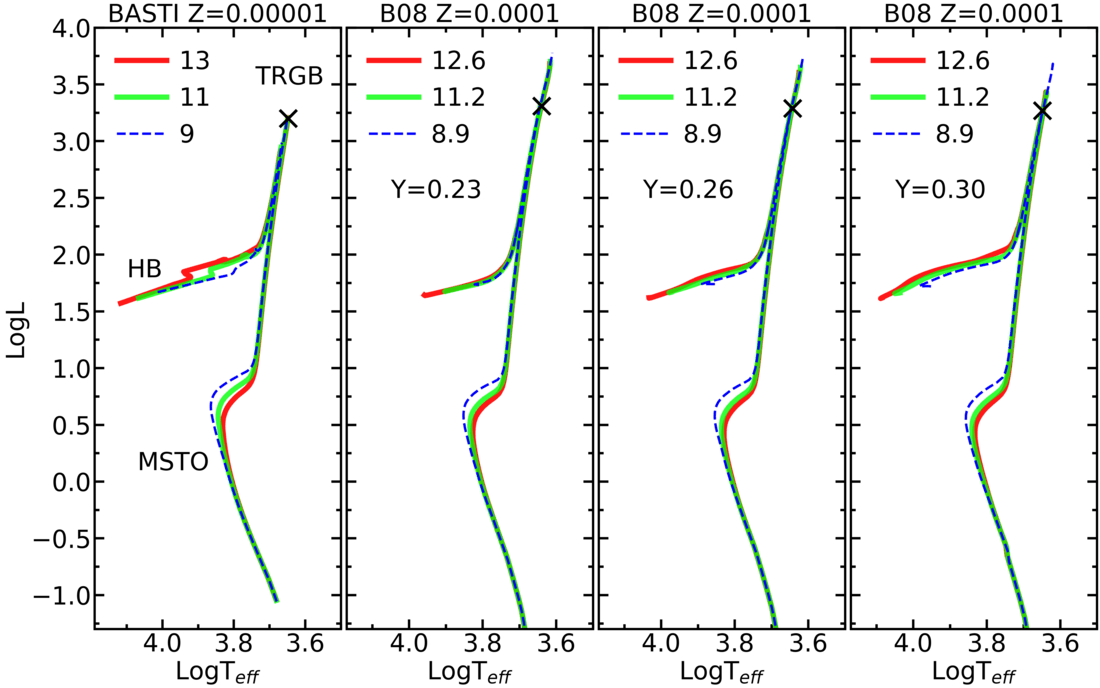} 
    \caption{Isochrones by \citet{P04} (BASTI) with the metallicity $\text{Z} = 0.00001$ and \citet{B08} (\citetalias{B08}) with the metallicity $\text{Z} = 0.0001$ and different age and Y. The X and Y axes, respectively, show the logarithm of the effective temperature in kelvins (logT$_\text{eff}$) and the luminosity logarithm in solar luminosities ($\rm log L$). Different colours represent the isochrones of different ages (in Gyr). The evolutionary stages of the MSTO and HB stars are marked in the left-hand panel.}
    \label{fig:2}
\end{figure*}

\begin{figure*}
    \centering
	  \includegraphics[width=0.65\textwidth]{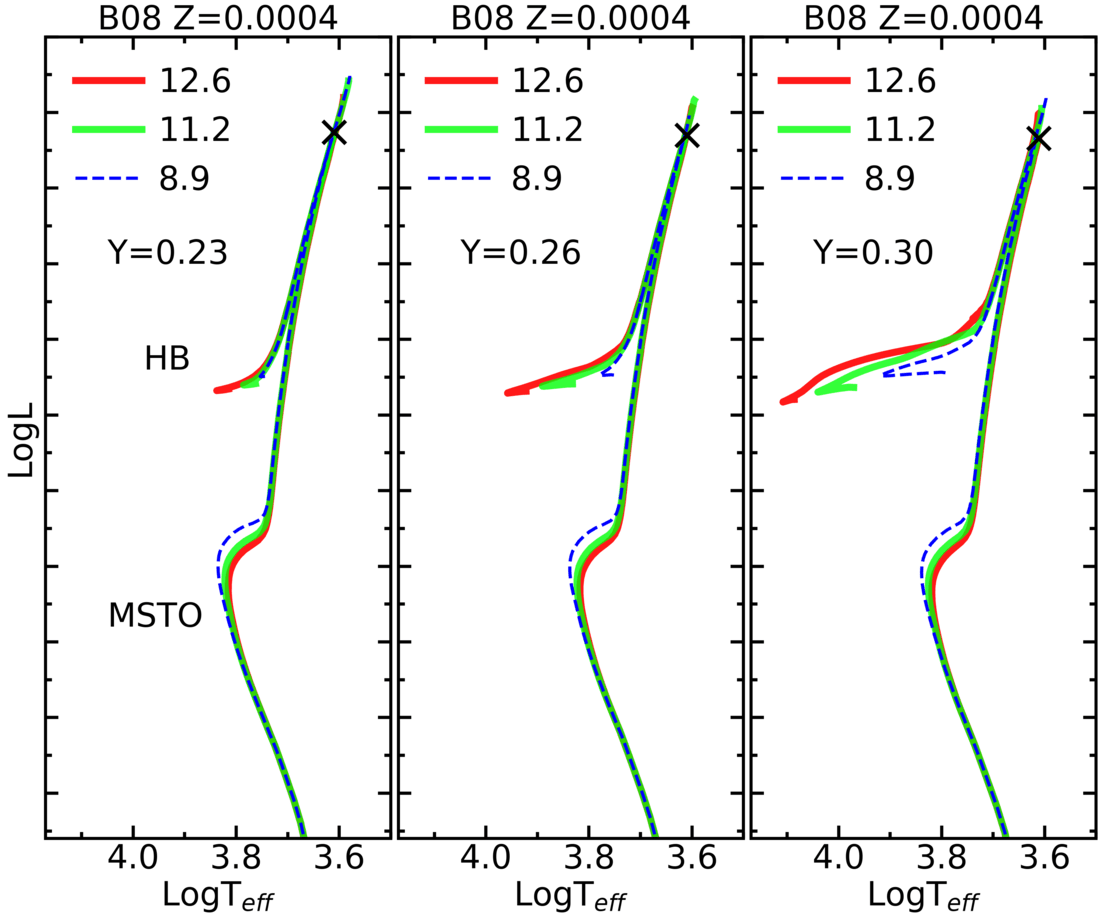}
    \caption{Isochrones by \citet{B08} (\citetalias{B08}) with the metallicity $\text{Z} = 0.0004$ and different age and Y. The X and Y axes, respectively, show the logarithm of the effective temperature in kelvins (logT$_\text{eff}$) and the luminosity logarithm in solar luminosities ($\rm log L$). Different colours represent the isochrones of different ages (in Gyr). The evolutionary stages of the MSTO and HB stars are marked in the left-hand panel.}
    \label{fig:A2}
\end{figure*}

\begin{figure*}
    \centering
	  \includegraphics[width=0.7\textwidth]{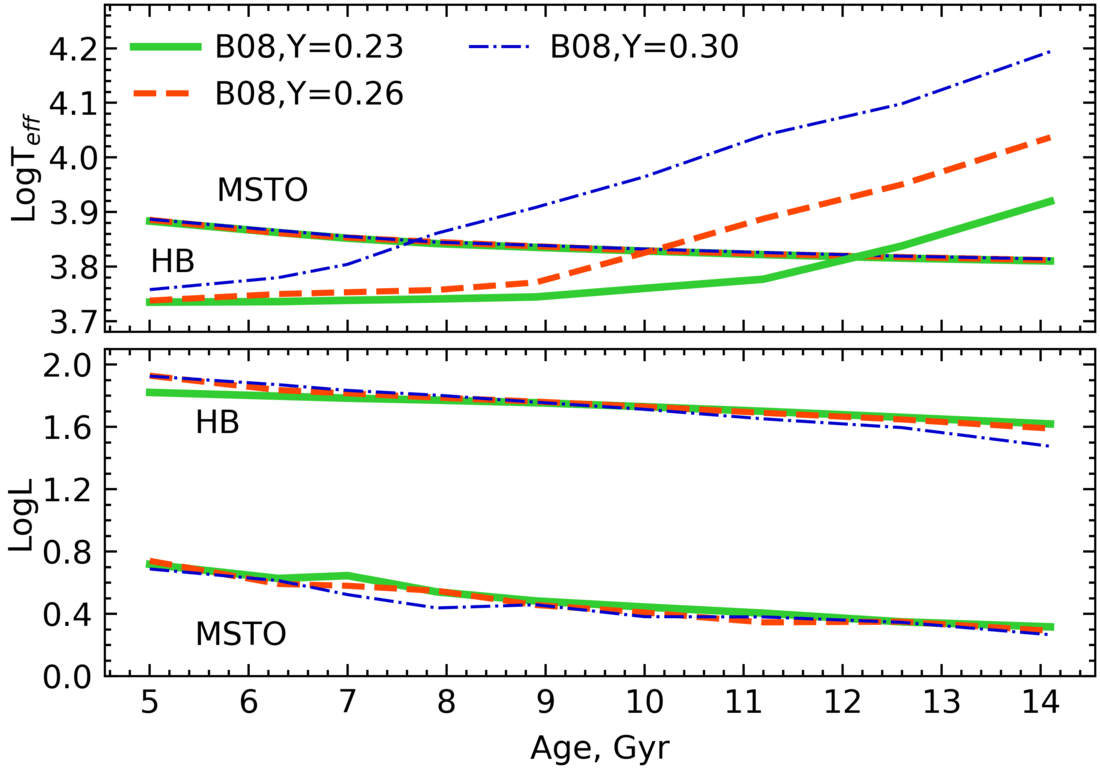}
    \caption{Effective temperatures (on the left) and luminosities (on the right) of the MSTO stars and the hottest HB stars depending on age. The \citetalias{B08} isochrones are used for Z=0.0004. The data for different isochrones are highlighted in colour as explained in the legend.}
    \label{fig:A3}
\end{figure*}

\begin{figure*}
    \centering
	  \includegraphics[width=0.9\textwidth]{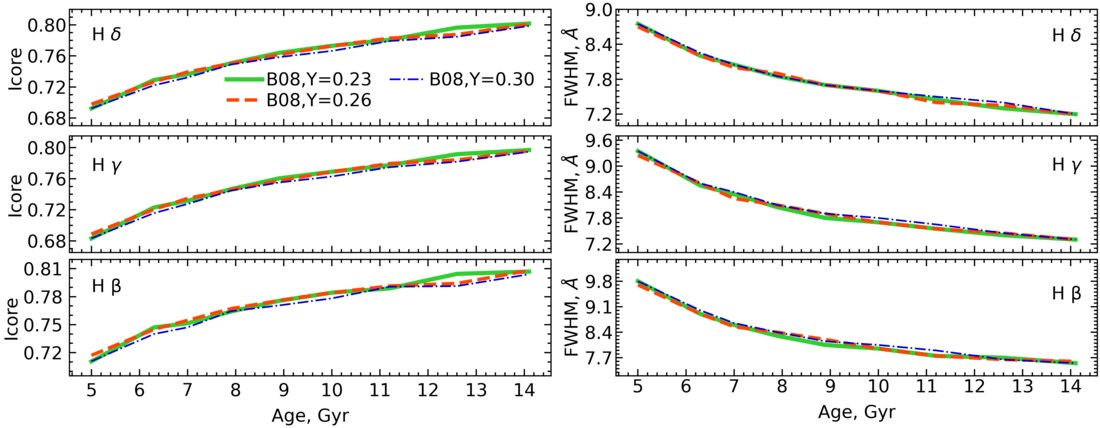} 
    \caption{Variation of I$_\text{core}$ and FWHM with age for three Balmer hydrogen lines in the synthetic IL spectra of GCs with the metallicity $\text{Z} = 0.0004$. The spectra were obtained using the \citetalias{B08} isochrones (the solid, dashed, and dash-dotted lines for $\text{Y} = 0.23$, 0.26, and 0.30, respectively) ignoring the HB stars.}
    \label{fig:A4}
\end{figure*}

 We illustrate the comparison of isochrones by \citetalias{B08} with various age and Y that we use for IL spectra modelling (Figs~\ref{fig:2}, \ref{fig:A2}). The IL spectrum of Ext~8 was approximated using the \citet{P04} isochrone with $\rm Z=0.00001$. The isochrones by \citet{P04} with $\rm Z=0.00001$ and various ages are displayed in Fig.~\ref{fig:2}. The dependence of $\rm log T_{eff}$ in K and $\rm LogL (L_{\sun})$ on the age for the metallicity $\rm Z=0.0004$ according to \citetalias{B08} are shown in Fig.~\ref{fig:A3}.
Figures~\ref{fig:A4} and \ref{fig:A4} demonstrate how the depth (I$_\text{core}$) and FWHM vary with age in case we choose the \citetalias{B08} isochrones with $\rm Z=0.0004$ (see Sec.\ref{sec:4} for details).


\clearpage

\begin{figure*}
    \centering
	  \includegraphics[width=0.9\textwidth]{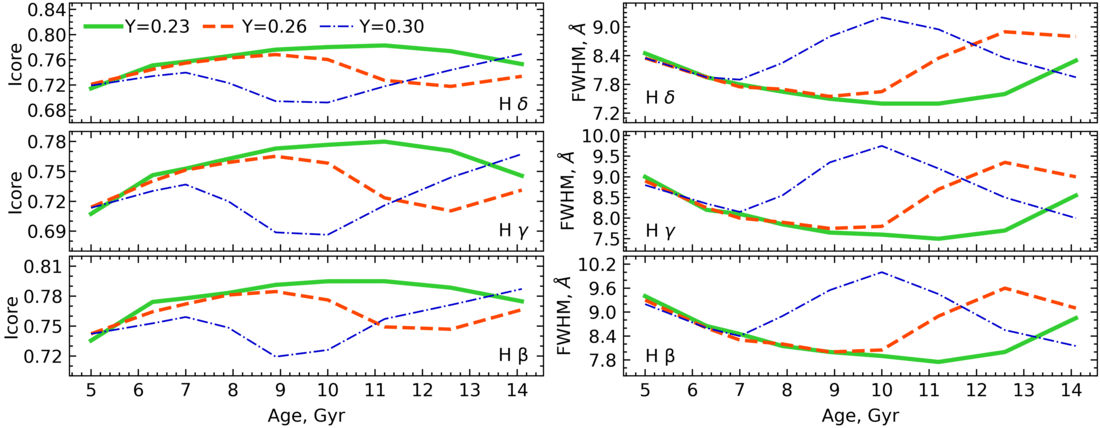} 
    \caption{Same as in the previous figure but with the HB stars included. }
    \label{fig:A5}
\end{figure*}

\clearpage
\section{How the observed and theoretical levels of the HB luminosity and of the effective temperature of the bluest part of the HB were determined for 35 Galactic GCs.} 
\label{app:3} 
 To compare the observed and theoretical levels of (1) the HB luminosity and (2) the effective temperature $\rm T_{eff}$ of the HB bluest part (see Sec.~\ref{sec:5_1} and Fig.~\ref{fig:T_L}), the following sequence of steps was performed. Hereafter, we will use the following abbreviations: "data"\ stands for "stellar photometry data for GCs in the Johnson-Cousins system by \citet{Piotto02} in the B and V bands or by \citet{Sar07} in the V and I bands"\ , "CMD"\  stands for "colour-magnitude diagram built using the data", and "theory"\  or "isochrone"\ stands for "theoretical isochrone of the stellar evolution by \citet{B08} selected by \citetalias{Sh20} using their analysis of the IL GC spectra of the Galactic GCs". It should be noted that in the analysis we use the absolute magnitudes corrected for extinction. 

(1) Determination of the HB theoretical luminosity and of the average $\rm log L(L{\sun})$ for the HB in the data: 

(i) The HB theoretical luminosity was determined in the middle of the HB according to $\rm T_{eff}$. The effective temperature of any point of the isochrone corresponds to the $\rm (V-I)_0$ and $\rm (B-V)_0$ theoretical colours at that point in the Johnson-Cousins system. 

(ii) In the data, a point with the same colour was selected on the HB. 

(iii) Stars belonging to the HB were selected in the data within $\pm$0.2~mag in colour (e.g., $\rm (V-I)_0$) and magnitude (e.g., $\rm M_V$) from this point. The CMD was considered for this purpose. If there were no stars on the HB within the specified limits, or their number was statistically insignificant, the colour and magnitude intervals to searches for stars in the middle part of the HB could be increased. Note, that on the real CMD, this point is located in the RR Lyrae gap between the red end of the blue HB (BHB) and the blue end of the red HB (RHB). If a GC had only BHB or RHB, then only the stars on the red end of BHB or only the stars on the blue end of RHB were selected. 

(iv) The HB luminosity levels $\rm M_V (HB)$ and $\rm M_I (HB)$ (or $\rm M_V (HB)$ and $\rm M_B (HB)$) and the corresponding standard deviations (STD) were estimated for the selected stars. 

(v) The resulting $\rm M_V (HB)$ and $\rm M_I (HB)$ (or $\rm M_V (HB)$ and $\rm M_B (HB)$) were superimposed on the dependence "absolute magnitude versus $\rm log L(L{\sun})$" for the HB in theory. For this purpose, isochrones of a given metallicity, corresponding to the metallicity of the theoretical isochrone (see step (i)), were used. Based on the correspondence between the observed $\rm M_V (HB)$ and $\rm M_I (HB)$ (or $\rm M_V (HB)$ and $\rm M_B (HB)$) and the theoretical luminosity, the average $\rm log L(L{\sun})$ for the HB in the data and the corresponding STD were determined. 

(2) Determination of the theoretical $\rm T_{eff}$ of the bluest HB point and of the average $\rm T_{eff}$ of the bluest HB part in the data: 

(i) On the isochrone, the bluest HB point was selected and the theoretical $\rm (V-I)_0$ or $\rm (B-V)_0$ and $\rm T_{eff}$ were determined for it from the isochrone. 

(ii) In the data, stars within the $\pm$0.2-mag range from the determined HB absolute magnitude level (in V and I or in B and V) on the bluest part of the HB in the $\pm$0.2-mag colour range were selected in order to determine the average colour $\rm (V-I)_0$ or $\rm (B-V)_0$ of the bluest HB stars. The CMD was considered for this purpose. If the number of stars within the specified magnitude range was statistically insignificant, the specified magnitude range could be increased. 

(iii) In the data, the average colour of the selected bluest HB stars and the corresponding STD were estimated. 

(iv) The average $\rm T_{eff}$ for the bluest HB stars, corresponding to the estimated average colour, was determined from the theoretical dependence "$\rm (V-I)_0$ or $\rm (B-V)_0$ versus $\rm T_{eff}$" for the HB. For this purpose, isochrones of a given metallicity (see step (i)) were used. 

After all, the differences between the observed and theoretical levels (data minus theory) of (1) the HB luminosity and (2) the effective temperature of the bluest HB part were estimated. 

\bsp	
\label{lastpage}
\end{document}